\documentclass[MS]{iitmdiss}
\usepackage{times}
\usepackage{t1enc}

\usepackage{graphicx}
\usepackage[colorlinks=true,linkcolor=blue]{hyperref} 
\usepackage{amsmath} 

\usepackage{algorithm2e}
\usepackage{algorithmic}
\usepackage{multirow}
\usepackage{subcaption,float}
\usepackage{graphicx}
\usepackage{etoolbox,array, blindtext}
\usepackage{booktabs}
\usepackage{hyperref}
\usepackage{color}
\usepackage[]{siunitx}
\usepackage{enumitem}

\setcitestyle{square}

\begin{document}


\title{Analysis of artifacts in EEG signals for building BCIs}

\author{Srihari Maruthachalam}

\date{February 2020}
\department{COMPUTER SCIENCE AND ENGINEERING}

\maketitle

\certificate

\vspace*{0.5in}

\noindent This is to certify that the thesis titled {\bf Analysis of artifacts in EEG signals for building BCIs}, submitted by {\bf Srihari Maruthachalam} (CS16S024), to the Indian Institute of Technology Madras, for the award of the degree of {\bf Master of Science (by Research)}, is a bona fide record of the research work done by him under my supervision.  The contents of this thesis, in full or in parts, have not been submitted to any other Institute or University for the award of any degree or diploma.

\vspace*{1.5in}

\begin{singlespacing}
\hspace*{-0.25in}
\parbox{4.5in}{
\noindent {\bf Prof. Hema A Murthy} \\
\noindent Research Guide \\ 
\noindent Professor \\
\noindent Department of Computer Science and Engineering\\
\noindent Indian Institute of Technology, Madras \\
\noindent Chennai, 600036 \\
} 

\end{singlespacing}
\vspace*{0.25in}
\noindent Place: Chennai\\
Date:

\acknowledgements

Firstly, I would like to thank my advisor Prof. Hema A Murthy, for trusting me and my ability to pursue research studies. She is one of the strongest, inspiring, and intellectual women, who consistently support me throughout my journey in Indian Institue of Technology Madras. She never missed to amuse me with her rational ideas, problem-solving ability, and immense writing skills. She always motivates and kept me on track. This thesis is possible because of her generous support. I am indeed fortunate to work with her.

I want to extend my gratitude to Prof. Mriganka Sur, Massachusetts Institute of Technology, for his wise ideas, suggestions, and support. With his guidance, my research studies in Neuroscience gained potential. It is my pleasure and honor to work with him. I thank Prof. Sriram Ganapathy, Indian Institute of Science, Bangalore, for ideating the artifacts classification work. He is down to earth and retains the energy to improvise the solution with his intellectual views. I thank him again.

I thank all my General Test Committee members, Prof. N.S. Narayanaswamy, Prof. V. Srinivasa Chakravarthy, and Prof. Mitesh Khapra. I extend my gratitude to all the faculty of the Department of Computer Science and Engineering for their enormous expertise, guidance, and assistance during my research studies.

I want to thank my friends and colleagues of the Speech and Music Technology (SMT) Lab. Special thanks to Mari Ganesh Kumar, Siddharth Agarwal, Rini Sharon, Karthik Pandia, and Saranya M S for their presence and support during my research studies. I extend my gratitude to my friends, Nauman Dawalatabad, Jom Kuriacose, Jeena J Prakash, Jilt Sebastian, P V Krishnaraj Sekhar, Arun Baby, Anju Leela Thomas, Anusha Prakash, Gayathri, Mahesh, Ashish Mishra, Vinoth Kumar, and Manish Jain.

Last but not least, I want to thank my family for their care and love. My eternal respect and love to my father, Maruthachalam, to my mother Kalavathy, and to my little sister, Abinayaa, for their support during my research studies.


\abstract

\noindent KEYWORDS: \hspace*{0.5em} \parbox[t]{4.4in}{Brain-Computer Interface, Electroencephalography, Text-To-Speech synthesis, Artifacts, Time warping techniques}

\vspace*{24pt}

\noindent Brain-Computer Interface (BCI) is an essential mechanism that interprets the human brain signal. It provides an assistive technology that enables persons with motor disabilities to communicate with the world and also empowers them to lead independent lives. The common BCI devices use Electroencephalography (EEG) electrical activity recorded from the scalp. EEG signals are noisy owing to the presence of many artifacts, namely, eye blink, head movement, and jaw movement. Such artifacts corrupt the EEG signal and make EEG analysis challenging. This issue is addressed by locating the artifacts and excluding the EEG segment from the analysis, which could lead to a loss of useful information. However, we propose a practical BCI that uses the artifacts which has a low signal to noise ratio.

The objective of our work is to classify different types of artifacts, namely eye blink, head nod, head turn, and jaw movements in the EEG signal. The occurrence of the artifacts is first located in the EEG signal. The located artifacts are then classified using linear time and dynamic time warping techniques. The located artifacts can be used by a person with a motor disability to control a smartphone. A speech synthesis application that uses eyeblinks in a single channel EEG system and jaw clinches in four channels EEG system are developed. Word prediction models are used for word completion, thus reducing the number of artifacts required.

\pagebreak


\begin{singlespace}
\tableofcontents
\thispagestyle{empty}

\listoftables
\addcontentsline{toc}{chapter}{LIST OF TABLES}
\listoffigures
\addcontentsline{toc}{chapter}{LIST OF FIGURES}
\end{singlespace}

\abbreviations

\noindent 
\begin{tabbing}
xxxxxxxxxxx \= xxxxxxxxxxxxxxxxxxxxxxxxxxxxxxxxxxxxxxxxxxxxxxxx \kill

\textbf{ALS}   \> Amyotrophic Lateral Sclerosis \\
\textbf{BCI}   \> Brain-Computer Interface \\
\textbf{BSS} \> Blind Source Separation \\
\textbf{CCA} \> Canonical Correlation Analysis \\
\textbf{CSF} \> Cerebrospinal Fluid \\
\textbf{CT} \> Computed Tomography \\
\textbf{DP} \> Dynamic Programming \\
\textbf{DTW}   \> Dynamic Time Warping \\
\textbf{ECG} \> Electrocardiogram \\
\textbf{ECoG} \> Electrocorticography \\
\textbf{EEG}   \> Electroencephalography \\
\textbf{EMG} \> Electromyography \\
\textbf{EOG} \> Electrooculography \\
\textbf{ERD}   \> Event Related Desynchronization \\
\textbf{ERP}   \> Event-Related Potentials \\
\textbf{ERS}   \> Event Related Synchronization \\
\textbf{FMRI} \> Functional Magnetic Resonance Imaging \\
\textbf{HCI} \> Human-Machine Interface \\
\textbf{HMM} \> Hidden Markov Model \\
\textbf{Hz} \> Hertz \\
\textbf{ICA} \> Independent Component Analysis \\
\textbf{IFCN} \> International Federation of Clinical Neurophysiology \\
\textbf{LTW}   \> Linear Time Warping \\
\textbf{MEG} \> Magnetoencephalography \\
\textbf{MRI} \> Magnetic Resonance Imaging \\
\textbf{ms} \> Milliseconds \\
\textbf{PCA} \> Principal Component Analysis \\
\textbf{PET} \> Positron Emission Tomography \\
\textbf{RNN}   \> Recurrent Neural Network \\
\textbf{SMR}   \> Sensorimotor Rhythms \\
\textbf{SNR}   \> Signal to Noise Ratio \\
\textbf{SSVEP}   \> Steady-State Visually Evoked Potentials \\
\textbf{TTS} \> Text To Speech \\
\end{tabbing}

\pagebreak


\chapter*{\centerline{NOTATION}}
\addcontentsline{toc}{chapter}{NOTATION}

\begin{tabbing}
xxxxxxxxxxx \= xxxxxxxxxxxxxxxxxxxxxxxxxxxxxxxxxxxxxxxxxxxxxxxx \kill
\textbf{$\delta$}  \> Delta wave in electroencephalogram signal\\
\textbf{$\theta$}  \> Theta wave in electroencephalogram signal\\
\textbf{$\alpha$}  \> Alpha wave in electroencephalogram signal\\
\textbf{$\beta$}  \> Beta wave in electroencephalogram signal\\
\textbf{$\gamma$}  \> Gamma wave in electroencephalogram signal\\
\textbf{$\Omega$}  \> Impedances of electroencephalography electrodes \\
\textbf{$\eta$}  \> The threshold used to detect artifact in EEG \\
\textbf{$\sigma$}  \> The standard deviation of the artifact signal \\
\textbf{$\mathcal{M}_n$}  \> $n^{th}$ sample in moving average of the electroencephalogram signal \\

\end{tabbing}

\pagebreak
\clearpage
\leavevmode\thispagestyle{empty}\newpage

\pagenumbering{arabic}


\chapter{Introduction}
\label{chap:intro}
Assistive technology enables disabled people to communicate and function more or less normally.  It could be a device or software. Further, assistive technology lessens the workload of caregivers. Assistive technology aids people from isolation, exclusion, and being locked-in; also, it helps in diminishing the influence of disease, limitation, and disability on a person, their family, and the community. One of the variants of assistive technologies is brain-computer interface (BCI), where neural electrical signals from the human brain are captured and interpreted as commands to control real-world devices.  In this thesis, we propose and develop an innovative artifact-based BCI.

\section{Electroencephalography (EEG)}

The human brain is the principal organ of the human nervous system. It manages most of the activities of the human body, namely, receiving, processing, integrating, and coordinating the information from the senses. Moreover, it plays a crucial role in decision making and sends instructions to the body. The brain is enclosed in, and shielded by, the skull of the head.

In recent times, a family of imaging techniques are being applied for examining brain functions of humans. These techniques include positron emission tomography (PET), magnetic resonance imaging (MRI), and computed tomography (CT). The measurements obtained from these techniques can produce exceptional spatial resolution of two or three-dimensional human brain images. On the other hand, electroencephalography (EEG), magnetoencephalography (MEG), and electrocorticography (ECoG) have an excellent temporal resolution for data processing compared with any imaging techniques.

Among the mentioned brain signal sensing techniques, EEG has a comparatively cheaper for recording. EEG is a non-invasive technique to record the electrical activity of the human brain from the scalp. EEG measures potential difference resulting from ionic current flow among the neurons of the brain. In the year 1929, Hans Berger measured the electrical activity over the human scalp and coined the term electroencephalogram for representing human brain electric potentials. It was also inferred that the EEG signal fluctuates over time and is related to prevailing cognitive states of subjects. Over the last few decades, researchers have actively studied the relationship between cognitive processes and the EEG signal.




EEG electrodes measure the relative electric potentials directly on the human scalp. When the brain nerve cells are activated, ionic potentials are produced. The electrodes placed over the scalp convert these ionic potentials into electrical potentials, which can be measured and stored. Since EEG is collected directly from the human scalp surface, the method can be applied frequently to healthy adults, children, and patients without risk and limitation. 

Since the EEG signal contains significant information corresponding to various physiological states of the brain, it is a commonly used tool for observing various neurological disorders.


\section{Brain-Computer Interfaces (BCI)}
A brain-computer interface (BCI) is a communication technique between an individual and external devices where the brain signals that are captured are translated. BCI is also referred to as a brain-machine interface (BMI) or direct neural interface (DNI) or a neural-control interface (NCI), a mind-machine interface (MMI). 

Currently, there are only a limited number of interfaces even for simple tasks, primarily because brain signals measured across sessions, across subjects are noisy and most challenging to interpret.  On the other hand, standard human-machine Interfaces (HCI) like the touchpad, buttons, gesture, and voice recognition are more robust. Nevertheless, BCIs are often propitious to the people, who lost control over most of the muscles, that otherwise could have been used as a medium of communication. For the people who are suffering due to total or partial locked-in syndrome, BCIs allow communicating with their surroundings. BCIs also improve a person's liberation and confidence.


\section{Artifacts}

Typically, the recorded neural EEG signals are in the range of microvolts, and it can be concealed by potentials generated from the non-cerebral source, which are called artifacts. So far, clinical applications and academic studies consider the presence of artifacts as a constant problem during the recording of brain activity. Among all the types of artifacts, electromyography (EMG) and electrooculography (EOG) artifacts are two notable contributors to physiological artifacts.


The most popular method for ocular artifact removal is based on linear combination and regression. In clinical and academic practice, the data affected with artifacts would be discarded in most situations, while it will lead to a significant information loss. There exists a variety of techniques for artifact separation and removal. By using a simple filtering technique, such as bandpass filters, the separation could be achieved in the frequency domain. Methods like blind source separation (BSS) \citep{joyce2004automatic}, canonical correlation analysis (CCA) \citep{lin2018real}, and independent component analysis (ICA) \citep{delorme2007enhanced}, \citep{levan2006system}, \citep{castellanos2006recovering}, \citep{wang2015removal}, \citep{tong2001removal} are currently employed in signal processing for eliminating the artifact. 

Minimal muscular movements like eye blink or jaw movement or head movement produce huge deflections in the EEG signal. Building a artifacts detection and classification model can help to augment a BCI. For example, associating the set of eyeblink patterns with commands or responses to the environment enables the people with partial locked-in syndrome to communicate.

\section{Problem statement}

The objective of this thesis is three-fold. 
\begin{itemize}
\setstretch{1.5}
\item The first objective is to analyze the artifacts generated from voluntary minimal motions of subjects in a controlled setup and study the properties and structure of the artifacts.
\item The second objective is to build a detection and classification model to classify the type of artifacts. It is achievable with the knowledge obtained from the first objective.
\item The third and final objective is to map the artifacts with the speller to enable real-world communication.
\end{itemize}
\section{Organization of the thesis}

The body of the thesis is explained in this section. This chapter provides an introduction to the area of brain-computer interface (BCI). Chapter~\ref{chap:intro} conveys a brief outline of the background material in terms of neuroscience, artifacts, Electroencephalography (EEG) devices, and existing techniques in BCI, which serve as background knowledge for Chapters ~\ref{chap:time} and ~\ref{chap:single}. Chapter~\ref{chap:time} proposes an efficient way to detect and classify the artifacts in EEG. Chapter~\ref{chap:single} proposes and elaborates the working of a BCI using single and four electrodes mobile EEG devices. Chapter~\ref{chap:sum} concludes the thesis.

\chapter{Background Material}
\label{chap:intro}
\section{Introduction}
Design and development of brain computer interfaces requires that we have a good understanding of the brain and its functions.   Using EEG signals for building BCIs requires an understanding of the EEG signal too.   A good understanding of EEG signals is required since the signals are weak, and are quite noisy.  Further, different lobes in the brain are responsible for different activities. It is equally important to study and understand the different time warping techniques, such as linear and dynamic time warping to classify the artifacts. In this thesis, we investigate brain electrical activity and artifacts from multiple electroencephalogram (EEG) recording devices for building an artifact-based BCI. 

The rest of the chapter is organized as follows. Sections~\ref{Sec:IntroHB} and~\ref{Sec:LobHB} provides an overview of the human brain and its anatomy, respectively. Section~\ref{Sec:BrainWav} describes the various frequency bands in electroencephalography (EEG). Sections~\ref{sec:artif}, ~\ref{Sec:TypArtiTool}, and ~\ref{Sec:TechClasArti} elaborates the artifacts, its types, and proposed techniques to classify the artifacts respectively. Section~\ref{Sec:EEGSetup} outlines the multiple electroencephalogram devices employed in the thesis. Sections~\ref{Sec:BCILit} and~\ref{Sec:AppBCI} discuss about existing BCI techniques and its applications respectively. The proposed BCI model is reported in Section~\ref{Sec:PropBCI}.

\section{Introduction to Human Brain}
\label{Sec:IntroHB}
The human brain is the principal organ of the human nervous system. It contains a volume of nerve tissues. The spinal cord makes up the central nervous system in the anterior end of the human body. The brain is also the center of learning, interpretation, comprehension, thinking, and language processing.

The brain contains the cerebellum, the brainstem, and the cerebrum. It manages most of the actions of the human body by integrating, processing, and organizing the information it collects from various sense organs and making logical decisions. The brain is included in and shielded with the skull bones of the head. The skull protects the brain, suspended in the cerebrospinal fluid, also called CSF. It is secluded from the bloodstream through the blood-brain barrier. Nevertheless, the brain is still sensitive to disease, damage, and infection. Damage can be induced by trauma or an injury or a lack of blood supply, referred to as a stroke. The brain is sensitive to degenerative diseases, such as Parkinson's disease, dementia, which includes Alzheimer's disease, and multiple sclerosis. Psychiatric medical conditions, including schizophrenia and clinical despair, are thought to be affiliated with brain dysfunctions. The pathological history of people with a brain impairment has given much insight within the purpose and capacity of every section of the brain. Moreover, brain study has evolved and grown with thoughtful philosophical, empirical, innovative experimental and theoretical aspects.

Many emerging techniques are employed to investigate and examine the brain. Medical imaging techniques such as functional neuroimaging or functional Magnetic Resonance Imaging (fMRI), computed tomography (CT), and positron emission tomography (PET). Imaging techniques are robust at spacial resolution. Medical monitoring techniques such as electroencephalography (EEG) and electrocorticography (ECoG) have a better temporal resolution. To build a practical BCI, the temporal resolution is more critical. These imaging and monitoring techniques have helped analyze how different parts of the human brain function to a certain extent. This understanding helps discriminate regions of the brain as specific lobes that can be associated with specific cognitive functions. In the following subsections, various parts of the human brain are discussed briefly.

\section{Lobes of the human brain}
\label{Sec:LobHB}
\begin{figure*}[!ht]
  \centering
\includegraphics[scale=0.4]{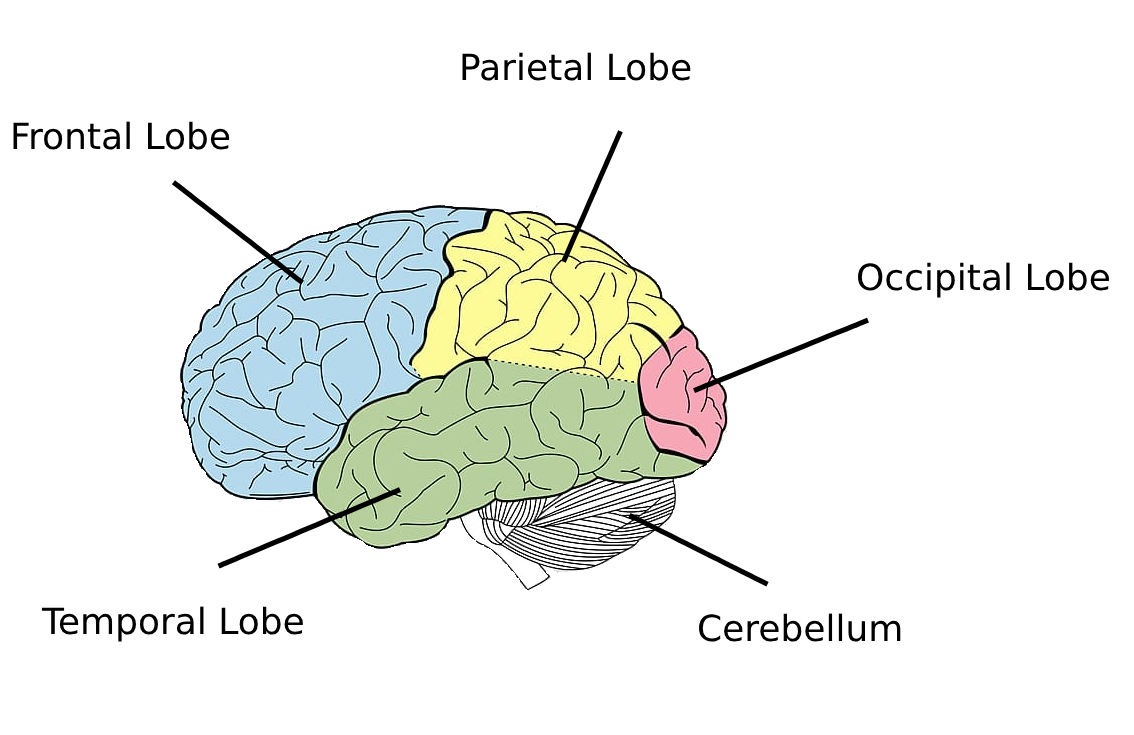}
  \caption{Lobes of the human brain}
  \label{Fig:Lobes}
\end{figure*}

The major part of the human brain is the telencephalon, which can be split into lobes. Respecting the anatomical distribution and various brain purposes, the cerebrum consists of six lobes of the brain \citep{ribas2010cerebral}. There are four significant lobes of the cerebral cortex in the human brain, and they are frontal, parietal, temporal, and occipital lobes, as shown in the Figure~\ref{Fig:Lobes}. Their locations and functions are briefly discussed below,

\vspace{-13mm}
\subsection{Frontal lobe}

The area at the front of the cerebral hemisphere is the frontal lobe, usually includes dopamine-delicate neurons, and it is responsible for short-time memory tasks, attention, motivation, and planning.

\subsection{Parietal lobe}

The parietal lobe is positioned over the occipital lobe and next to the frontal lobe. The sensory information from various modalities is aggregated in this lobe, including spatial sense and tactile sense of the skin. Besides, some regions in the parietal lobe perform a crucial role in language processing.

\subsection{Temporal lobe}

The region underneath the lateral gap of cerebral hemispheres is the temporal lobe. It is incorporated with visual remembrances, emotion association, and language comprehension.

\subsection{Occipital lobe}

The occipital lobe is the visual processing core of the brain. It is also responsible for diverse tasks, such as motion perception, visuospatial processing, and color differentiation.

\section{Brain waves}
\label{Sec:BrainWav}
A continuous recording of the electrical activity of the brain for brain scientists and clinical specialists uses two fundamental parameters, namely amplitude, and frequency, which describe the EEG signal. Some EEG patterns are considered reliable for visual examination. In general, there are five standard brain waves classified by different frequency bands, and these rhythms are named by Greek letters individually as $\delta$ (delta), $\theta$ (theta), $\alpha$ (alpha), $\beta$ (beta) and $\gamma$ (gamma). Berger discovered the alpha and the beta waves in the year 1929 \citep{millett2001hans}. Jasper and Andrews coined the gamma wave in the year 1938 \citep{jasper1938brain}. Delta and theta waves were introduced by \citep{walter1953living}. Delta activity belongs to EEG activity in the range of 0.5-3 Hz. It is often affiliated with EEG synchronized sleep in people. In the first few years of a human infant's life, the dominant frequency is the delta wave. Theta activity can be observed by a low-frequency range of 3-8 Hz. \citep{schacter1977eeg} shows that the theta wave is related to two psychological phenomena. The first one is a low level of attentiveness and sleep loss states. The second one is in perceptual processing and problem-solving. The theta band is also responsible for active and effective processing. Alpha activity happens across the posterior areas of the human head and happens when the individuals are awake and relaxed. The alpha wave consists of comparatively high voltage, which usually is smaller than 50 $\si{\micro} V$, across the occipital regions in the range of 8 to 13 Hz. The presence of alpha activity is associated with physical relaxation with closed eyes \citep{schomer2012niedermeyer}. The general conscious rhythm of the brain is the beta wave, and it corresponds to active attention, active thinking, and solving problems in healthy adults. Usually, the beta wave has low-voltage changes with a range of 13 to 30 Hz. A higher frequency spectrum from 30 to 70 Hz or more with lower voltage changes is the gamma wave.

\section{Artifacts}
\label{sec:artif}
As per the International Federation of Clinical Neurophysiology (IFCN), an artifact is described as any potential difference due to an extra-cerebral origin, which is shown in the EEG signal \citep{kane2017revised}. The impact on the EEG signals by artifact is seen as a well-recognized challenge for experimental and clinical electroencephalography. For a long period of time, the artifact in EEG activity was regarded as an obscuration \citep{klass1995continuing}. Sometimes, artifacts may have similar parameters in frequency, rhythmicity, and recurrence compared to the recorded electric potentials of the cerebral source. So it would become considerably challenging to figure out the contrast between artefactual and cerebral electrical activity \citep{brittenham1974recognition}. Because the EEG signals are always in the range of milli to microvolts, they can be easily masked by artifacts. Typically, based on their source, the artifact can be classified into two categories, physiological and non-physiological artifacts. The origins of physiological artifacts are the non-neural movements of the subjects, such as muscle activities or eye movement. While the non-neural artifact originates from outside of the human body, such as environment or equipment, improper attachment of the electrodes in a controlled environment is one of the common reasons for this technical artifact \citep{anderer1999artifact}. In the brain-computer interface (BCI) study, electromyography (EMG) and electrooculography (EOG) artifacts are the most commonly identified sources of physiological artifacts.

Numerous research works have reported that EMG and EOG activities could influence the neurological aspects involved in a BCI system. For example, early target-related EMG based artifact is manifested during the initial stages of BCI training and is discussed in \citep{mcfarland2005brain}.

\section{Types of artifacts measurement interfaces}
\label{Sec:TypArtiTool}
In this section, some popular types of artifacts measurement interfaces are summarized and argued why EEG is adequate to measure any artifact.

\subsection{Electromyogram (EMG)}

Muscle tissue contraction can move various parts of the human body. These muscles are classified as smooth, skeletal, and cardiac muscles. Electromyogram (EMG) exhibits the electrical activities of skeletal muscles, where the action potentials originate between the muscles and the nervous systems. Generally, a noninvasive electrode or needle electrode could be employed to measure myoelectric signals. In research studies, with the aid of surface electrodes on the skin, the surface EMG is examined. The recording of surface EMG could be affected by various types of noise, such as electrode motion artifact, which may limit signal quality \citep{sornmo2005bioelectrical}. Cranial EMG has numerous characteristics that are accountable for the harmful effects on the EEG activity. First, it was reported that the EMG has a wide frequency range, which is from 0 to 200 Hz \citep{goncharova2003emg}, which indicates that EMG activity influences all the EEG bands, including delta, alpha, and beta bands. The EMG is topographically distributed in the whole human body. So when the energy of muscle contraction raises, EMG influences and affects the whole scalp.

\vspace{-15mm}
\subsection{Electroculogram (EOG)}

Electroculogram is the electrical activity produced by eye movement, which has a significant effect on EEG recording. The potential difference between the cornea and retina can be altered by eye movement, which exists not only in the awake state but also during sleep. The strength of EOG mostly relies on the electrode close to the eyes and the direction of eye movement (horizontal or vertical). Besides, the potential difference could be influenced by blinking, which is generated by the muscle movement of the eyelid. This type of ocular activity produces a different waveform, and it may only occur during the awake periods. The blinking artifact has a high frequency, and the amplitude is significantly more in the frontal electrodes. For artifact processing, EOG signals can be measured by using reference electrodes placed near the eye in practice. As a common type of artifact, EOG can present severe complications in EEG analysis due to the proximity to the brain \citep{sornmo2005bioelectrical}.

\subsection{Electrocardiogram (ECG)}

The electrical activity of the heart can be measured by electrocardiogram (ECG). Regular heartbeats can be described by a repetitive, commonly occurring wave pattern, which is suitable to expose the appearance of the ECG artifact. The amplitude of cardiac activity is weaker on the scalp compared to the EEG signals on the scalp.  However, if ECG is visible in the background EEG signals, the cardio artifact may be overlooked as brain activity. Like the eye-related artifacts, the ECG can be measured separately by placing several reference electrodes alongside cerebral activity.

\vspace{-13mm}
\subsection{Technologic artifacts}

Being an uncontrolled variation in the experimental settings, artifacts can be introduced at the time of data acquisition, primarily the equipment that connects the subjects and EEG instruments. The common artifacts are the electrode, the electrode-scalp interface, the jack plug, and the input cable. These types of experimental artifacts are nearly impossible to avoid or reduce. One potential cause of artifact is the electrode cable, which connects the electrode with the data acquisition equipment. Owing to the inadequate shielding in practice, the electric current running from nearby powerlines or electrical appliances produce electromagnetic fields.

Consequently, 50/60 Hz powerline interference modifies the EEG signals. The movement of electrodes can modify the DC connection potential and provide an artifact called an electrode-pop artifact. This technical artifact may happen not only in the EEG signals but also in any bioelectric signals recorded on the body surface. The electrode-pop artifact has frequently acted as an abrupt shift in the baseline level. A massive movement by subject could change the location of the electrode on the scalp. The changes in distance between the skin and the recording electrode result in signal distortion. Moreover, the conduction capacity among the electrodes would be altered by the movement of the recording tools with respect to the underlying skin \citep{sweeney2010simple}. It was explained that the artifact produced by the electrode-scalp interface hugely depends upon the skin condition of subject, as well as the kind of conductive gel applied. 

%

\subsection{EEG for artifacts}

The muscle tissue contraction based artifacts are measured with EMG. However, \citep{goncharova2003emg} shown that the EEG is sufficient to capture the muscle tissue contraction based artifacts. EOG measures the electrical activity produced by eye movements. Nevertheless, EEG is capable of measuring the same \citep{wang2015removal}, \citep{zeng2014removal}. Similarly, ECG can be captured with the aid of EEG \citep{tong2001removal}. Technological artifacts can be detected and eliminated from the EEG signal with signal processing techniques \citep{tatum2011artifact}. It is evident that EEG is sufficient to measure most of the artifacts produced. 


\section{Techniques to classify artifacts}
\label{Sec:TechClasArti}
Simple time-warping techniques are used to detect the type of artifacts. The EEG signal pattern corresponding to that of the artifact is quite strong and evident.   As a first attempt, we attempt to match the EEG signal potentials across different realizations of the artifact.   As two different realizations of the artifact need not be of the same duration and amplitude, we employ different types of warping techniques.

In order to compare two time-varying signals, the incorporation of elastic distance measures is more justified in contrast with lock-step distance measures. Elastic distance measures allow one-to-many or many-to-one or one-to-none point matching. This phenomenon makes it viable for elastic distance measures to warp in time and be more robust in estimating the similarities and dissimilarities. Two of the crucial elastic distance measures in existence are Linear Time Warping \citep{ltw2017} and Dynamic Time Warping techniques \citep{berndt1994using}, which aid in finding the alignment between two time series. These two techniques are discussed in the following subsections.

\subsection{Linear Time Warping}

Linear Time warping (LTW) works with the underlying principle of linear interpolation. Given any two known points in the coordinates $(a_0, b_0)$ and $(a_1, b_1)$, the linear interpolant is the straight line between the points. For any value $a$ in the interval $(a_0, a_1)$, the corresponding value $b$ can be estimated by, 
\begin{equation}
b = b_0 \left(1-\frac{a-a_0}{a_1-a_0}\right) + b_1 \left(\frac{a-a_0}{a_1-a_0}\right)
\end{equation}

\begin{figure*}[!ht]
\begin{minipage}[h]{1.0\linewidth}
 \centering
        \includegraphics[scale=0.4]{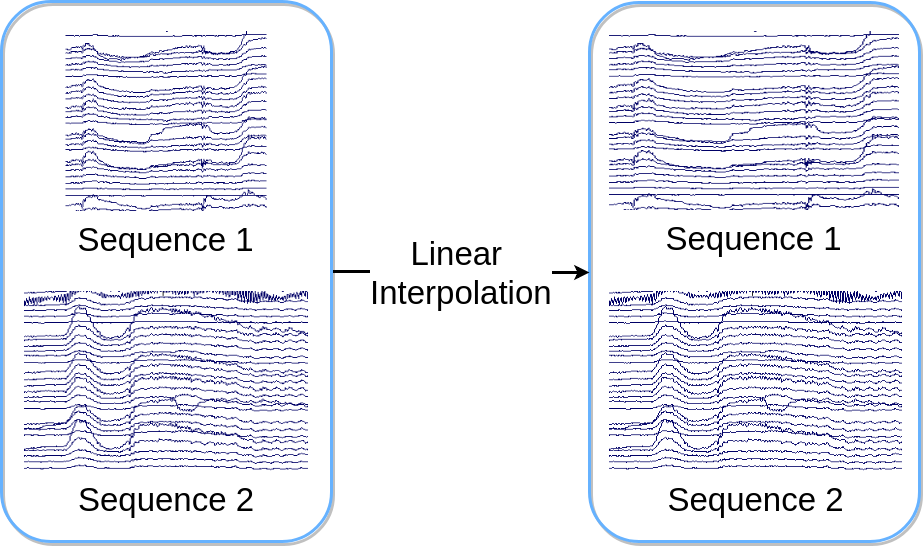}
  \caption{Illustration of linear interpolation of sequences}
  \label{Fig:LTW}
      \end{minipage}
\end{figure*}

LTW is a technique that is applied to determine the alignment between two temporal series. The two temporal series are interpolated to a particular length, and the Euclidean distance between the series is computed. 

The estimated distance provides us similarities and dissimilarities between the two series. Consider two time series X and Y, of lengths n and m respectively, and let $n > m$,
$$ X = x_1, x_2, x_3, ..., x_n $$
$$ Y = y_1, y_2, y_3, ..., y_m $$
The interpolation of the series in LTW can be performed in three different ways. First, the interpolation of the smaller sequence to that of the larger sequence, that is, making $ Y = y_1, y_2, y_3, ..., y_n $; second, the interpolation of the larger sequence to that of the smaller sequence, that is, making $ X = x_1, x_2, x_3, ..., x_m $, however, this interpolation technique is not recommended as there might be some information loss; third, the interpolation of both of the series to a predefined length, say k, making $ X = x_1, x_2, x_3, ..., x_k $ and $ Y = y_1, y_2, y_3, ..., y_k $. In the literature, first and third interpolation techniques are widely used.

Once the two series are interpolated to a fixed-length series as shown in the Figure~\ref{Fig:LTW} and the Euclidean distance metric is employed to estimate the similarity between the series.

\subsection{Dynamic Time Warping}

Dynamic Time warping (DTW) is an algorithm, applied to determine the alignment between two temporal series (a template and a query). It was widely used in old speech recognition applications. Speech recognition typically implies the translation of spoken utterances to textual words. In the DTW based speech recognition, audio information is transformed into templates. The template is matched with every template by implementing some constraints. The most suitable match template perpetually has the least distance measure from the input query. Nowadays, the application of DTW is no more confined to speech recognition; as a matter of fact, it can be practiced with any temporal data, which can be interpreted in a linear series. Originally proposed DTW algorithm has quadratic space and time complexities and is explained in \citep{keogh2001derivative}, \citep{senin2008dynamic}, \citep{ratanamahatana2004everything}, \citep{salvador2007toward}, \citep{berndt1994using} literature. 

As discussed earlier, DTW can be practiced on any temporal data which can be presented in a linear series, which can vary in speed and time. To comprehend DTW, let us consider two time series X and Y of the length n and m respectively,
$$ X = x_1, x_2, x_3, ..., x_n $$
$$ Y = y_1, y_2, y_3, ..., y_m $$
DTW adopts a dynamic programming strategy to obtain the alignment between the two time series, which aligns the time series based on optimally minimized distance. Dynamic Programming also called as DP, is a robust methodology that breaks the massive problem into smaller sub-problems. The output of smaller sub-problems is determined and then aggregated to compute the solution of the original problem. Due to this characteristic, it is additionally appreciated as a divide and conquer strategy. That is, remembering the past output as the output of the sub-problem, which contributes to answering the original problem. It can be accomplished by employing a couple of different methods, and they are top-down and bottom-up. In the top-down method, the problem is resolved by dividing it down into smaller sub-problems. Sub-problems are resolved independently, and the outcome of every sub-problem is collected and stored. This process ultimately contributed to an overall resolution. At the same time, the bottom-up strategy of DTW, where the outcome of sub-problem is applied to resolve the given original problem progressively. For illustration, in DTW, the sub-problems D(i, j) is solved first as given in Equation \ref{eq:DTW_3}, and applied those progressively to compute the solution for D(n, m).

The initial step to estimate DTW alignment between two time series is to form an n-by-m cost matrix where every $(i^{th}, j^{th})$ element corresponds to distance estimated between $x_i$ and $y_j$. Distance can be measured in by applying various distance metrics, say, simple Manhattan difference $d(x_i, y_j) = \left| x_i - y_i \right|$, Euclidian distance or squared distance $d(x_i, y_j) = \left( x_i - y_i \right)^2$, or any other distance metric functions. \citep{akila2013slope} comprises of the review of various distance functions which can be used in DTW. Based on aggregative distance for every path in the cost matrix, the best-suited match between time series can be obtained by using Euclidian measure, which is one of the most popular methods in distance metric computation.
\begin{equation} \label{eq:DTW_1}
D(i, 1) = D(i - 1,1) + d(i, 1)
\end{equation}
\begin{equation} \label{eq:DTW_2}
D(1, j) = D(1, j - 1) + d(1, j)
\end{equation}
\begin{equation} \label{eq:DTW_3}
D(i, j) = d(i, j) + min[D(i - 1, j), D(i - 1, j - 1), D(i, j - 1)]
\end{equation}
%
%
%
%

Dynamic programming formulation can be asymmetric and symmetric as well. In asymmetric formulation, one of the places around the diagonal, that is, D(i - 1, j) or D(i, j - 1) is skipped or provided higher weight. Nevertheless, Equation \ref{eq:DTW_3} can be considered as a symmetric formulation, considering both points around the diagonal of the current point, are given the same weights.

Research analyses report that symmetric formulation proffers more stable returns as contrasted to asymmetric in speech recognition \citep{berndt1994using}. A DP formulation, for illustration, Equation \ref{eq:DTW_3}, in the given context, provides the aggregative measure for every point by considering the sum of distance measure with the least of aggregative measures of the neighboring diagonal points. It supplies the global cost matrix D with filling the first row and first column of the matrix, in the following manner as given in Equations \ref{eq:DTW_1}, \ref{eq:DTW_2}, and \ref{eq:DTW_3} by initializing D(0,0) = 0

Once the global cost matrix has been filled with aggregated distances, the following step is obtaining the warping path between the two time series through the cost matrix. Variants of DTW are proposed in the literature.   The pros and cons of these variants are discussed in Section~\ref{sec:dtw}.


An optimum path consists of a series of continuous matrix cells passing through the cumulative cost matrix, which defines the alignment between two time series is called warping path. The warping path can be determined by implementing dynamic programming formulation, which is also called as step patterns. The formulation is provided in Equation~\ref{eq:DTW_3}. The warp path exploration begins from D(n, m) and backtracks by the evaluation of all neighboring cells from left, down, diagonally through the bottom left. If any of those neighboring cells hold the least values, and they are appended to the origin of the warping path continuously till D(1,1) is reached.

\begin{figure*}[!ht]
  \centering
\includegraphics[scale=0.4]{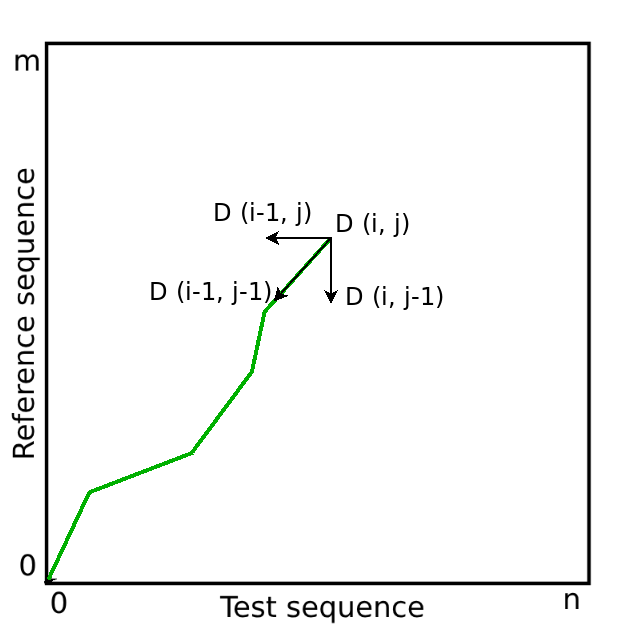}
  \caption{Dynamic Time Warping and warping path}
  \label{Fig:DTWIntro}
\end{figure*}

Figure~\ref{Fig:DTWIntro} shows the execution of DTW to determine the warping path between the template and the query time series. DTW algorithm begins from position D(0,0) and propagates till the point at the highest position, which is D(n, m). At every position D(i, j), the aggregated distance for each position is determined by taking the total sum of distance d(i, j) with the smallest aggregated distance of all the successor positions as given in Equation \ref{eq:DTW_3}. A symmetric step pattern, where all the neighboring positions engage on an equal basis. One DP formulation is assessed for all the positions, and the cost matrix is formed with aggregated distance measures. It is forthwith feasible to obtain the optimal warping path by backtracking from the position(n, m), Figure~\ref{Fig:DTWIntro} presents the potential warping path which can be achieved after backtracking the cost matrix. From the observations across the optimal warping path taken from the cost matrix, several enhancements have been proposed in the literature and introduced to as constraints.

\subsubsection{Boundary constraint}

Boundary condition concentrates on the start and end of the warping path. It affirms that the first position of the warping point needs to be $w_1$ = (1,1), and the last position should be $w_k$ = (n, m), where n and m denote the duration of query and template time series respectively. The warping paths that do not satisfy boundary conditions are usually marked as incorrect paths.

\begin{figure*}[!ht]
  \centering
\includegraphics[scale=0.4]{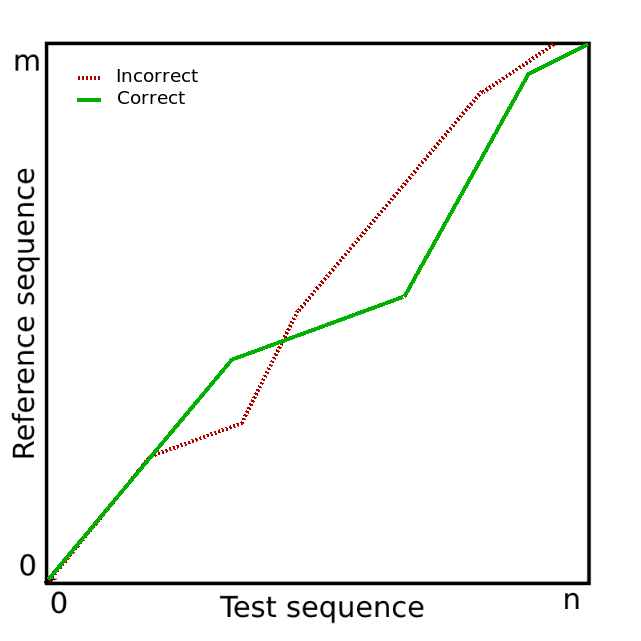}
  \caption{Boundary constraint in Dynamic Time Warping}
  \label{Fig:DTWBound}
\end{figure*}

Graphical illustration of boundary constraint is presented in Figure~\ref{Fig:DTWBound}, where the solid line originates from the first position of the cost matrix and ends on the last position $w_k$ = (n, m). At the same time, the dashed line begins from the primary position but ends at $w_k$ = (n - 1, m), which breaks the boundary constraint. Therefore it can be termed as an incorrect warping path.

\subsubsection{Continuity constraint}

Continuity constraint takes care of maintaining a valid warping path. In other words, it makes sure the participation of every position in both query and template series. It can also be explained as; every cell in the cost matrix should be the restricted to its adjacent cells, or the previous position of any point in the cost matrix must be $(i_k - 1, j_k ), (i_k , j_k - 1), (i_k - 1, j_k - 1)$.

\begin{figure*}[!ht]
  \centering
\includegraphics[scale=0.4]{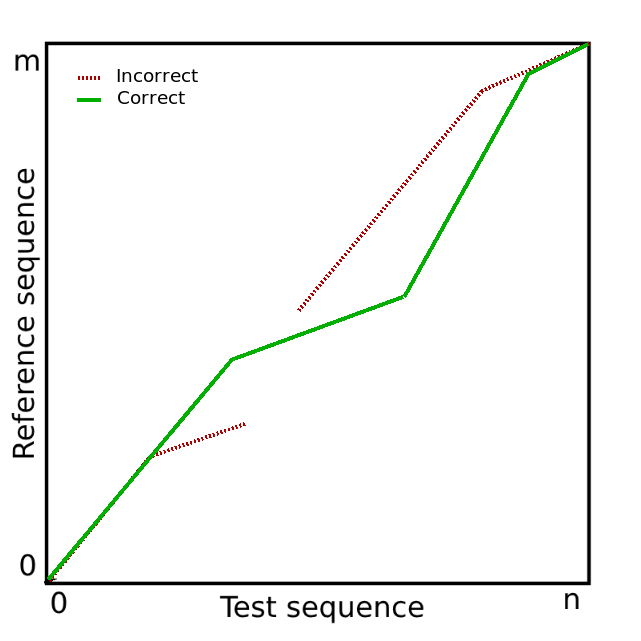}
  \caption{Continuity constraint in Dynamic Time Warping}
  \label{Fig:DTWCon}
\end{figure*}

Figure~\ref{Fig:DTWCon} explains the continuity constraint, and it can be observed that the path represented by the dashed line is satisfying the boundary constraints but opposing the continuity constraint. An optimal path should obey all the constraints.

\subsubsection{Monotonicity constraint}

\begin{figure*}[!ht]
  \centering
\includegraphics[scale=0.4]{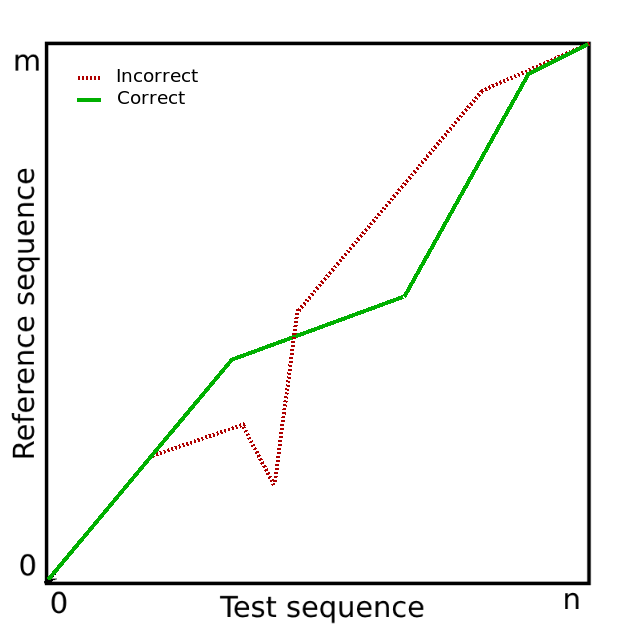}
  \caption{Monotonicity constraint in Dynamic Time Warping}
  \label{Fig:DTWMon}
\end{figure*}

A proper warping path needs to be continuous and have valid boundary points and obligated to be monotonic. Monotonicity constraint demands that the points in the warping path should have an increasing or nondecreasing trend. It can be said as a warping path cannot decrease in time, it can be straight or can be increasing, that is, $i_k \ge i_k - 1$ and $j_k \ge j_k - 1$ for all the steps.

The warping path outlined by the dashed line in the Figure~\ref{Fig:DTWMon}, meet all other DTW constraint, however, has decreased in the time for a while, which is sufficient to ignore them. An accurate warping path needs to satisfy all three above discussed constraints that are, continuity constraint, boundary constraint, and monotonicity constraint.

\vspace{-13mm}
\section{Electroencephalography setups}
\label{Sec:EEGSetup}
In the thesis, we carried out experiments on three different EEG data collection equipments. In this section, the data acquation apparatus are explained in detail.

\subsection{Single electrode EEG setup}
\label{sec:SEEG}
\begin{figure*}[!ht]
  \centering
\includegraphics[scale=0.4]{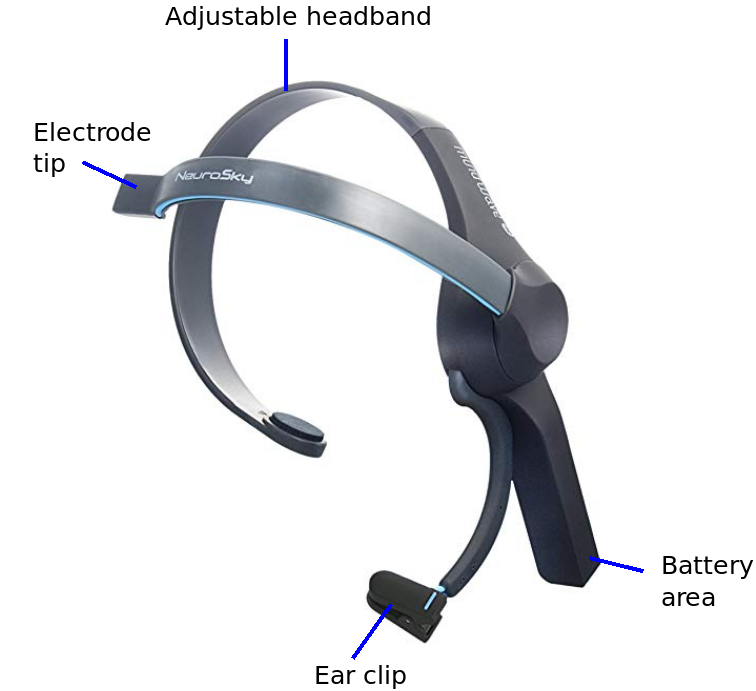}
  \caption{Single electrode EEG by Mindwave Mobile}
  \label{Fig:SEEG}
\end{figure*}

NeuroSky Mindwave Mobile \citep{neurosky2019} consists of eight components, which are ear arm, ear clip, battery area, adjustable headband, power switch, electrode arm, electrode tip, and think gear disk as shown in the Figure~\ref{Fig:SEEG}. The Mindwave mobile needs a AAA battery that runs for 8 hours continuously. Bluetooth 2.1 is employed with 1.0V minimum expected voltage, 10mA power consumption, and 10m connectivity range. The electrode on the forehead identifies the electrical signal from the frontal lobe of the human brain. The second electrode is an ear clip, which is employed as a ground to sift out the electrical noise and the ambient noises. The Mindwave mobile carefully estimates and outputs the EEG power spectrums such as alpha, beta, theta, delta, and gamma waves. Moreover, it transcends the traditional wet electrodes with the conduction gel or saline for extended time EEG measurement \citep{lin2010novel}. 

The thesis utilizes NeuroSky Mindwave Mobile for several reasons. First, the project strives to contribute to a low-cost practice, which can be employed by everyone. Second, its signal is transmitted in digitized form through Bluetooth \citep{blondet2013wearable}.

One principal lack is the exactness of the EEG signal obtained by NeuroSky Mindwave Mobile because the NeuroSky Mindwave Mobile Mobile possesses only one electrode, FP1, which is in the frontal lobe. Another viable issue is comfort. Subjects report it is uncomfortable to wear for a long time.

\subsection{Four electrodes EEG setup}
\label{sec:FEEG}
Muse \citep{muse2019} is a brainwave-sensing headband designed by InteraXon, Inc. that is marketed to customers as a system to improve attention, focus, and control. The headband weighs 57 grams and includes four recording electrodes, which are two electrodes on the forehead and two electrodes on the back of both ears. They are FP1 (left forehead), FP2 (right forehead), TP9 (back of the left ear), and TP10 (back of the right ear). It does not need the application of conductive gel during the recording. Muse shows EEG and transfers the data to a mobile or a computer through Bluetooth. Wireless connection is carried through Bluetooth 2.1 + EDR, and data is sampled and transmitted at 500 Hz, electrodes are manufactured with Silver for FP1 and FP2, Conductive silicone-rubber for TP9 and TP10. The battery is capable of running for 5 hours continuously. Muse measures the behavior of five brainwaves: delta, alpha, beta, gamma, and theta, and shows the participant's brainwave activity as active, neutral, or calm. 
\begin{figure*}[!ht]
  \centering
\includegraphics[scale=0.4]{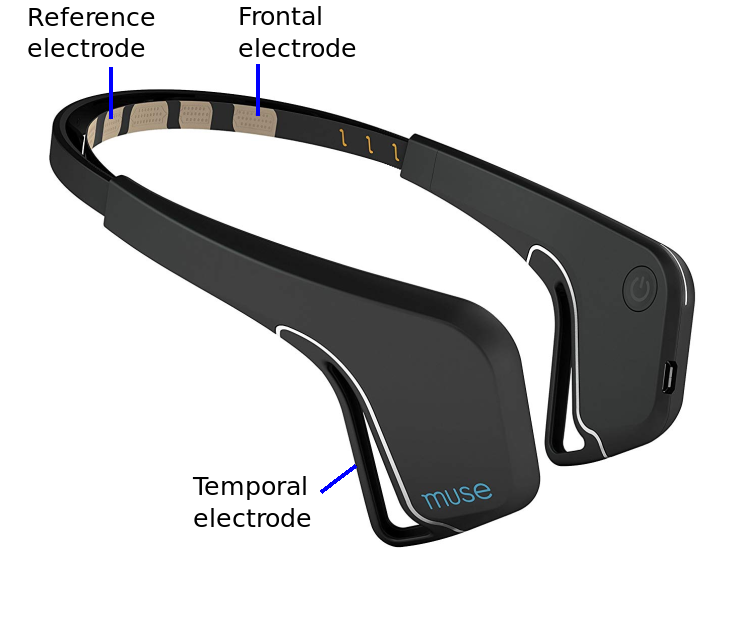}
  \caption{Four electrodes EEG by Muse}
  \label{Fig:FEEG}
\end{figure*}

One of the setbacks of this device is not capable of changing the wearer's brainwave patterns, reading his or her thoughts, or helping the wearer to move things using only thoughts.

\subsection{128 electrodes EEG setup}

\begin{figure*}[!ht]
  \centering
\includegraphics[scale=0.3]{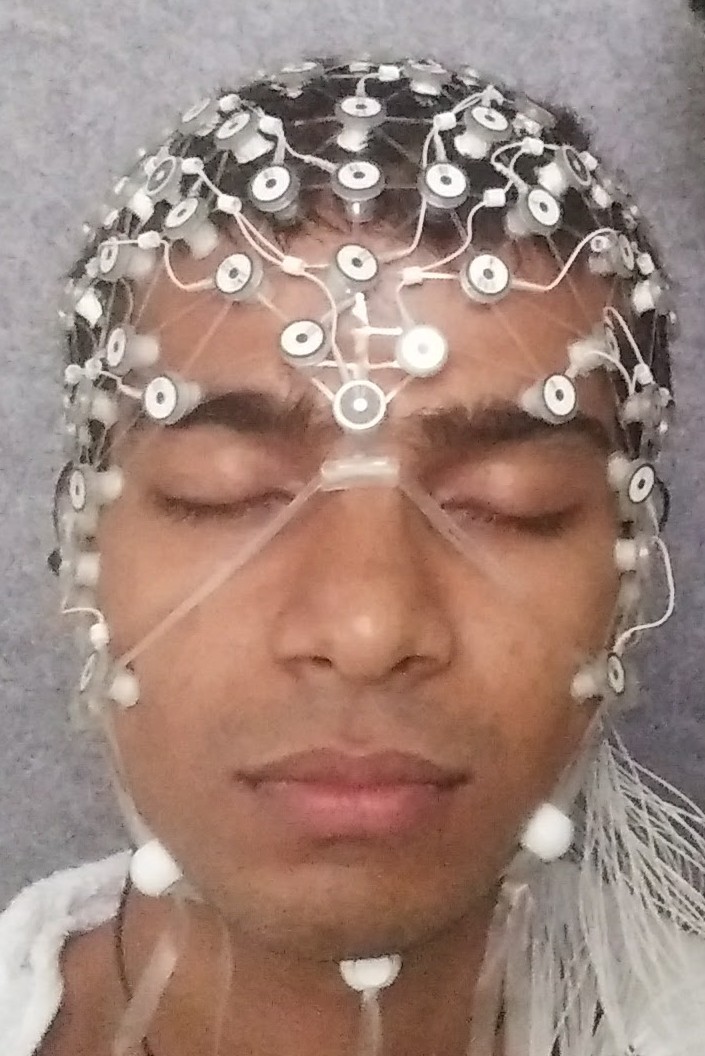}
  \caption{128 electrodes EEG by EGI}
  \label{Fig:128EEG}
\end{figure*}

EGI (now acquired by Philips) \citep{egi2018web},  \citep{egi2019} Geodesic Sensor Net (GSN) is devised to obtain dense-array EEG data utilizing a Geodesic EEG System (GES) and shown in the Figure~\ref{Fig:128EEG}. The electrodes that make up the GSN 400 are bound into a geodesic construction employing long-lasting polyurethane elastomer fibers that form the tension lines of various icosahedra. While the GSN 400 is expanded over a subject's scalp, the electrodes produce electrical contact with the scalp and are continuously held in place. The tension is equally spread across all electrodes, giving a comfortable array that can be worn for a few hours. The scalp point of the electrode pedestal is expanded, making the pedestal to raise and level the scalp hair as it is pushed against the head. This setup provides a self-seating effect, so that the pedestal sponge, amidst its load of electrolyte, is included below the hair, immediately upon the scalp. All electrodes, including reference and solitary common, are held in the Net's construction. This design makes the application of the Nets a quick process. 

Standard application times are shorter than 10 minutes with impedances in the 10 to 50 $k\Omega$ range. The GSN 400 is designed for use with HydroCel Saline electrolyte, which is EGI's standard potassium chloride saline and surfactant solution. Enclosing each electrode pellet is a sponge which, during the execution of the GSN 400, is soaked with HydroCel Saline electrolyte. The wet sponge swells from the one edge of the pedestal, and the lead cable rises from the other end into a small space in the pedestal caplet. The electrode array is attached to an amplifier. The amplifier measures the EEG signals that are picked up by the electrode array and samples them at milliseconds. The scalp should be washed and kept dry while recording.

\section{Brain-Computer Interface models}
\label{Sec:BCILit}
A Brain-Computer Interface (BCI) is a systematic way that maps central nervous system signals and interprets the data to output, suitable for a machine to employ as an input signal. Alternatively, a BCI is a communication conduit that provides for primary control of a machine using one's thoughts. Although it is usually considered that the BCI as being a pretty new area, the term and idea "Brain-Computer Interface" was coined by Jacques J. Vidal in 1973 earlier \citep{vidal1973toward}. The primary objective of a BCI is to map brain signals, investigate and understand the determined data, and to transmute the interpretation into operations \citep{dornhege2007toward}, \citep{wolpaw2012brain}. 

An illustration for a BCI setup could be a subject who is resting in the face of a pinball machine by the responsibility to control the flippers of the pinball device by his/her thoughts \citep{tangermann2008playing}. The subject's brain activity could be mapped using EEG. During that, the subject wears an EEG net with electrodes, which are measuring the electrical activity on the scalp of the subject, produced by the brain activities beneath the scalp, in the brain. The computer accepts continuous data from all EEG electrodes and interprets the data. The computer can render the subject's imagined left or right hand action into a signal that classifies "right" or "left" whenever the subject imagines the corresponding hand movement. The computer is attached to the pinball device. Whenever it receives the "right" or "left" signals, the right and left flipper flips. The subject can operate the pinball machine using BCI. 

Development of a BCI is dependent on three crucial tasks, namely, i) data acquisition, ii) analysis of brain signal, and iii) converting the same to an appropriate action as requested by the subject. In the given example, feedback is a flipper response. However, the feedback can occur in many shapes or colors or both. It can be the moving flipper of the pinball device or could be a prosthesis \citep{muller2007control}, virtual keyboards \citep{farwell1988talking}, \citep{maruthachalam2018brain} or yet the steering wheel of an automobile \citep{zhao2009eeg}. The critical characteristic of feedback is that it transmutes the output signal into some desirable action.

\subsection{Event-related potentials (ERPs)}

Event-related responses are elicited responses despite the characteristics of the stimulus. To a particular event, such as an auditory, visual stimulus, and motor action, there exist stimuli-specific brain responses. Unlike natural EEG, Event-related potentials (ERPs) needs averaging methods across trials to elicit useful information. Typically, the initial step of the ERP method to identify the time of each event and mark them as epochs. Every epoch has a definite duration. After repeating the stimulus several times, the epochs are averaged across trials. The averaged event-related response can be obtained at both individual and group levels. The classification of the ERP element can show the polarity (negative or positive deflection), scalp distribution, and timing. ERPs have a precise temporal resolution, and the human brain activity can be measured on a range of tens of milliseconds, and many aspects of perception and attention in operation \citep{woodman2010brief}.  One of the common challenges in ERP analysis is the misinterpretation of the relationship between the visible peaks and the underlying elements. A set of rules was proposed by \citep{luck2005ten} to avoid misunderstanding the relationship between the visible peaks and the underlying elements.


\subsection{Steady state visually evoked potentials (SSVEPs)}
The SSVEPs are brain responses that are accurately synchronized with flickering visual stimuli. When the subject pays attention to a visual stimulus, that can be a blinking led, for example, a blinking LED, the response produced in the occipital lobe captures quasi-sinusoidal rhythms that match the frequency of the blinking sinusoid. Consequently, frequency elements with the corresponding value of the stimulus frequency and its harmonics are detectable using spectral analysis techniques. The strength of the frequency element depends directly on how focused the subject is on the stimulus. The advantage of this method is that the SSVEP response is strong across diverse subjects and usages that can provide relatively fast communication. Furthermore, this model does not need a significant amount of training. A subject can produce a strong SSVEP response even without proper training  \citep{graimann2010brain}. However, the disadvantage of the SSVEP is that the subject needs to maintain a constant gaze at the blinks on the screen, which can lead to fatigue.

\subsection{Attention Based BCI}

Attention-based BCI serves by conferring different subjects with different stimuli. Such stimuli can be auditory \citep{klobassa2009toward}, \citep{schreuder2010new}, visual \citep{farwell1988talking}, \citep{citi2008p300}, \citep{allison2010bci}, or tactile \citep{muller2006steady}. Each stimulus is correlated with a particular action, like the movement of a portion of a prosthesis, or the choice of a character from the alphabet. By concentrating his/her attention on the desired stimulus and neglecting the others, the subject provides the brain signal patterns that the BCI system requires to "interpret" his/her intention and perform the desired action \citep{dornhege2007toward}, \citep{wolpaw2012brain}. From all the attention-based BCIs, the visual attention-based BCI is widely used. In this case, two various brain signal patterns are used, and they are event-related potentials (ERP) and steady-state visually evoked potentials (SSVEP). In ERP-based BCI, the stimuli are displayed successively and for a short amount of time. The span of a presentation is normally a few milliseconds and the time within two stimuli around 100ms. When the particular stimulus is presented, and the subject focuses his/her attention on the stimulus, the subject generates a brain signal that is distinct from when he/she is not performing a focused activity.

The most prominent pattern is an event-related positivity in the region of centro-parietal areas around 300 to 500 milliseconds after the display of the stimulus. This phenomenon is called a P300 or P400 or P500 based on the time taken after the stimulus. The ERP-based BCI system can identify the ERP and therefore distinguish between the "targets," which are the stimuli the subject was attending, and the "non-targets." In contrast to ERP based BCI, where the stimuli are displayed successively, in SSVEP based BCI, stimuli are displayed continuously, all at the very same time, and flickering with various frequencies between 6-30Hz. The subject keeps his target by concentrating on a particular stimulus. The flickering frequency of the stimulus produces SSVEP with the corresponding frequency in the visual cortex region. For illustration, if the subject focuses on a stimulus flickering with a frequency of 23Hz, the SSVEP will flicker with a frequency of 23Hz too. The BCI system can identify the stimulus the subject was attending to by analyzing the frequency of the SSVEP with the frequencies of the stimuli. 

Visual attention based BCI serves reliably over various subjects. A speller based model can be built with visual attention based BCI. Nevertheless, visual models might need the subject to have control over his/her eye gaze in order to work accurately, which is not always addressed for patients, particularly people suffering from locked-in syndrome. A possible solution could be gaze independent visual spellers \citep{treder2010c}. Another pretty solid problem with visual spellers is that several people despise the constant flickering on the display screen and find it extremely exhausting to practice over an extended time.

\subsection{Motor Imagery-Based BCI}

When a muscle in the human body is voluntarily moved, there are variations in human brain electrical activity in the motor cortex and sensorimotor regions. This is referred to as sensorimotor rhythm (SMR). These variations are comparatively localized, following the homuncular structure of the cortical area \citep{woolsey1979localization}. The fall of oscillations is called event-related desynchronization (ERD) and usually arises during the movement or the preparation of movement. The rise of oscillations is called event-related synchronization (ERS) and usually arises after the movement or rest \citep{pfurtscheller1999event}. Imagining such physical movements provide substantially similar ERD or ERS signal patterns as the real movements would, and it is discussed in \citep{pfurtscheller1997motor}. Motor imagery-based BCI serves by expecting the subject to imagine the physical movement of particular limbs, like gripping with the left or right hand or relocating the feet and measuring the ERD/ERS signal patterns across the corresponding cortical areas. Matching a specific imagined limb movement with a distinct action that gets performed helps in building motor imagery-based BCI. Although the motor imagery model resembles extremely natural opposed to the attention-based BCI, the number of muscles a person can move and the number of complicated actions each muscle can perform makes motor imagery based BCI challenging to build.

Opposed to attention-based BCI, motor imagery based BCI has a tremendous error rate. About 15\% to 30\% of subjects are not capable of obtaining control employing motor imagery based BCI without proper training \citep{blankertz2010neurophysiological}. Another critical limitation for healthy subjects is that one cannot practice motor imagery-based BCI while doing something else. Since every movement the subject performs creates ERD or ERS signal patterns, the subject has to avoid any movement in order to practice motor-imagery based BCI accurately.

\section{Applications for BCI}
\label{Sec:AppBCI}

The most critical application or the purpose of BCI is assistive technology for severely physically impaired people. The standard instance, found in numerous BCI publications, are subjects who suffer from Amyotrophic Lateral Sclerosis (ALS) \citep{charcot1874sclerose} a neurodegenerative medical condition that gradually paralyzes the victim until he/she is utterly locked-in in his/her paralyzed form. In the latter stage, the patient is conscious, however incompetent to twitch even a single muscle in his/her body (locked-in syndrome). With the help of BCI, we can implement spellers \citep{farwell1988talking}, \citep{birbaumer1999spelling} that enable victims to communicate to the outer world without any or with minimal muscular movement. Along with communication, mobility is another critical application for BCI. Stroke victims might use BCI to visualize the current state of the brain in order to determine how to suppress undesired signal patterns \citep{daly2008brain}. Alternatively, subjects with spinal cord lesions can use BCI to command a wheelchair \citep{galan2008brain}, a telepresence device \citep{escolano2010telepresence}, or even a prosthesis \citep{birbaumer2007brain}. Nevertheless, the low information transfer speed of a non-invasive BCI is the principal barrier for composite applications using BCI. For illustration, a hand prosthesis that is managed through BCI will typically not enable the subject to achieve low-level movements like specific thumb control, but solely a small collection of high level controls like "open hand" and "grasp." A subject who is capable of twitching some muscles in his/her body voluntarily will often be quicker and more reliable using those muscles to manage a device than practicing BCI \citep{mak2009clinical}. Apart from assistive technology, BCI has further found importance in various fields \citep{blankertz2010berlin}. Researches have assessed how BCI technology can be applied as calibration equipment to measure mental states like attention \citep{schubert2008parieto}, \citep{haufe2011eeg}, or workload \citep{kohlmorgen2007improving}, \citep{muller2008machine}, \citep{venthur2010novel} in order to predict and perhaps prevent human errors in critical circumstances. BCI technology is employed for quality evaluation by measuring the subconscious comprehension of noise in visual or auditory signals \citep{porbadnigk2010using}, \citep{porbadnigk2011revealing}. BCI can further be employed in entertainment and gaming \citep{krepki2007berlin}, \citep{nijholt2009turning}, and the gaming business has commenced producing games which are "mind-controlled."

\section{Proposed Artifacts based BCI}
\label{Sec:PropBCI}
As we discussed in Section~\ref{sec:artif}, a minimal muscular artifact can influence the EEG signal to a large extent. So, most of the studies are aimed to reduce or eliminate the artifacts in the EEG signal. Taking advantage of the influence of artifacts on the EEG signal, an attempt is made to detect artifacts in the EEG signal, and build a BCI using the artifact. In the following chapters, simple, but effective threshold-based artifact detection and robust time-warping techniques to classify the artifacts are proposed; with the aid of detection and classification models, effective and easy to use BCIs are built and demonstrated.

\section*{Summary}

In this chapter, we discussed the basic physiology of the human brain, electrical activity in the cerebral region of the brain, and methodologies to captured the electrical activity. We reviewed the various bands in the electrical activity of the brain. We discussed the artifacts and their types. We briefly studied the linear and dynamic time warping techniques. We explained the fundamentals of Brain-Computer Interfacing and its literature. We briefed the importance of the proposed artifacts-based Brain-Computer Interfacing.

\chapter{Time Warping solutions for classifications of artifacts}
\label{chap:time}
\section{Introduction}
In the previous chapter, we briefly discussed the proposed artifacts classification based Brain-Computer Interface (BCI). In this chapter, we study the effect of minimal muscular artifacts such as a head turn, a head nod, a jaw movement, and an eye-blink and their data collection process. It is observed that the artifacts have temporal signatures. We take the benefit of these temporal signatures and develop methods to detect and classify them.


Section~\ref{sec:LitArtif} deals with past work on analysis of artifacts. Section~\ref{Sec:DataColl} describes the experiment setup and data collection techniques. Preprocessing of the acquired EEG data is discussed in Section~\ref{Sec:Prep128EEG}.  The detection of artifacts in the preprocessed EEG signal is detailed in Section~\ref{sec:thres}. Sections~\ref{sec:classArti}, \ref{sec:ltw}, and \ref{sec:dtw} discuss the classification of artifacts with Linear and Dynamic Time Warping techniques. Section~\ref{sec:det} explains the methodologies to detect and classify the artifacts in a continuous EEG signal.

\section{Related Work in Artifacts}
\label{sec:LitArtif}
EEG signals comprise some valuable information about brain \citep{anderson1995determining}. In addition to this valuable erudition, EEG also apprehends artifacts, which are assumed to be undesirable electrical potentials that arise from non-cerebral sources. Artifacts are addressed by first finding and eliminating the EEG signal segment from the study \citep{whitton1978spectral}. This approach could drive to a loss of valuable data in the EEG signal. An alternative strategy would be to diminish the impact of the artifact in the EEG signal. Occasionally, the artifacts are employed to develop BCIs \citep{maruthachalam2018brain}, \citep{ma2014using}, Prof. Hawking used cheek twitches to convey messages to the world \citep{hawking2012}. In \citep{nolan2010faster}, an automated threshold on the amplitude of the EEG signal based artifacts rejection procedure was introduced. Wavelet, Kurtosis, and Renyi's entropy-based examination to identify the artifacts in the EEG signal were presented in \citep{inuso2007brain}.  In \citep{rohalova2001detection}, techniques for EEG artifacts detection based on Kalman filter autoregressive model and radial basis function neural networks were introduced. All the procedures discussed above are agnostic to the class of artifacts in the EEG signal. Our work not only discovers the artifact; it also classifies the type of artifacts. Independent Component Analysis (ICA) in EEG signals has been broadly practiced for artifacts elimination and can be found in \citep{jung1998extended}, \citep{mammone2012automatic}, \citep{winkler2011automatic}. An effort has been made to identify and eliminate blink artifacts and eyeball movement from EEG signal data applying blind component separation in \citep{joyce2004automatic}.  In \citep{jiang2007automatic}, detection and elimination of heartbeat artifact from EEG signal data utilizing wavelet analysis was introduced.  In \citep{brunner1996muscle}, muscular artifacts were detected using spectral analysis was proposed. In this chapter, we propose a novel thresholding method to detect the artifacts in the EEG signal.  Following the artifacts detections, time warping algorithms such as Linear Time Warping (LTW) and Dynamic Time Warping (DTW) are adopted to classify the artifacts.

\section{EEG Data Acquisition} 
\label{Sec:DataColl}
Electroencephalography (EEG) signals were acquired using saline-based 128 electrodes EEG array setup built by Electrical Geodesics, Inc (EGI) \citep{egi2018}. Clean and dried EEG net was carefully immersed and soaked in Potassium Chloride electrolyte for five minutes. After five minutes, the EEG net was carefully placed and covered the entire scalp of the subjects. The EEG net was connected to an amplifier. The output of the amplifier was fed to a computer machine to quantize and store the output EEG signal. EEG signals were acquired at a sampling rate of 250 Hertz. The impedances of all the electrodes were maintained below $50k\Omega$ during the experiment. Throughout the experiment, the subjects were asked to keep their eyes closed unless explicitly stated in the instruction. The experiments were conducted in a controlled setting where explicit instructions were given to the subjects. Subjects were expected to make the muscular action corresponding to a given artifact instruction voluntarily. Subjects were requested to perform four kinds of muscular movements as follows, mouth open and close, head nod, eye open and close, and head turn left and right.

\begin{figure*}[!ht]
  \centering
\includegraphics[scale=0.5]{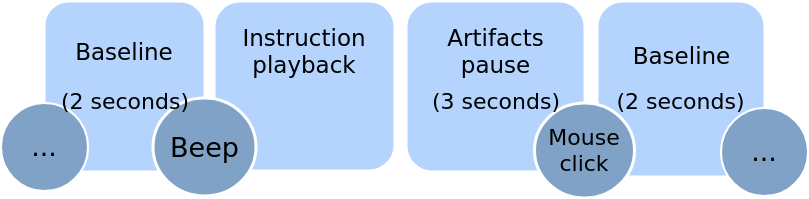}
  \caption{Illustrative timeline of an artifact acquisition trial}
  \label{Fig:ExpSetup}
\end{figure*}

Instructions were given to the subject with a loudspeaker, in a random order, to perform the muscular movement. Before every trial, the subject was given two seconds time interval to rest. The EEG signal corresponding to the resting time is considered as the baseline or resting state EEG signal. The timeline for data acquisition throughout a trial, with the resting state, is shown in Figure~\ref{Fig:ExpSetup}. Subjects were asked to pay attention to the instruction entirely and make a muscular movement accordingly. Three seconds of time was given to the subjects to do the muscular actions after listening to the instruction playback. Following the muscular movement, the subjects were requested to produce a mouse click.  This mouse click is to guarantee that they indeed made the muscular movement. 
These controlled experiments were conducted with nine subjects. Out of the nine subjects, four subjects appeared for another session after six months. All the subjects were acquainted with the purpose and scope of the research, and signed consent was also received to acquire their EEG data\footnote{The Ethics Committee of Indian Institute of Technology Madras approved the study.} \footnote{The EEG setup is supported by the project, CSE/12-13/132/UPFX/HEMA.}. An average of 22 trials was obtained in each session.

\begin{figure*}[!ht]
\begin{minipage}[h]{1.0\linewidth}
 \centering
        \includegraphics[scale=0.4]{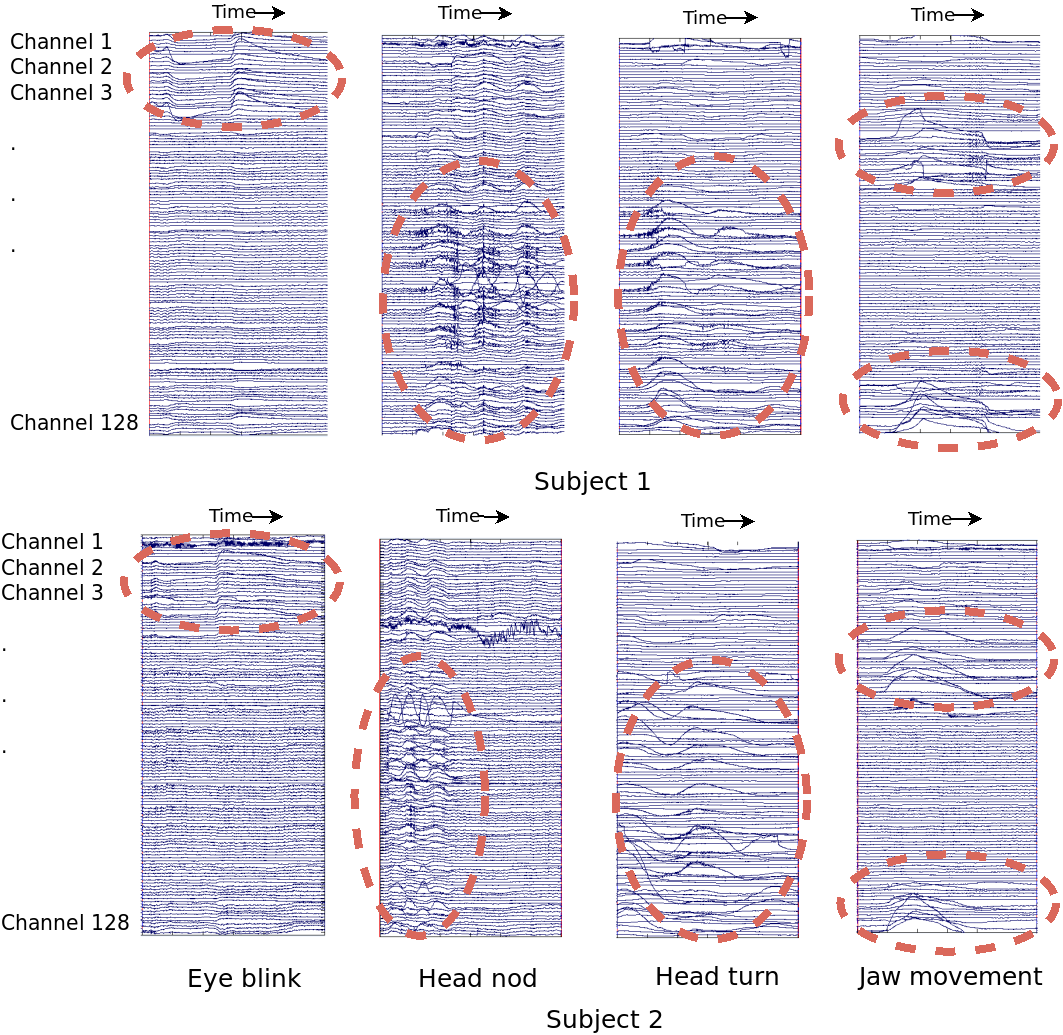}
  \caption{EEG signals of different artifacts of two subjects}
  \label{Fig:Example}
      \end{minipage}
\end{figure*}

EEG artifacts of two separate subjects are presented in Figure~\ref{Fig:Example}, and from the figure, it can be observed that the temporal signatures of various artifacts are different even across subjects\footnote{Supplementary plots for artifacts of multiple subjects are accessible in \href{http://bit.ly/EEGPlot}{http://bit.ly/EEGPlot}.}. 

\section{Preprocessing of EEG Data} 
\label{Sec:Prep128EEG}
\begin{figure*}[!ht]
  \centering
\includegraphics[scale=0.45]{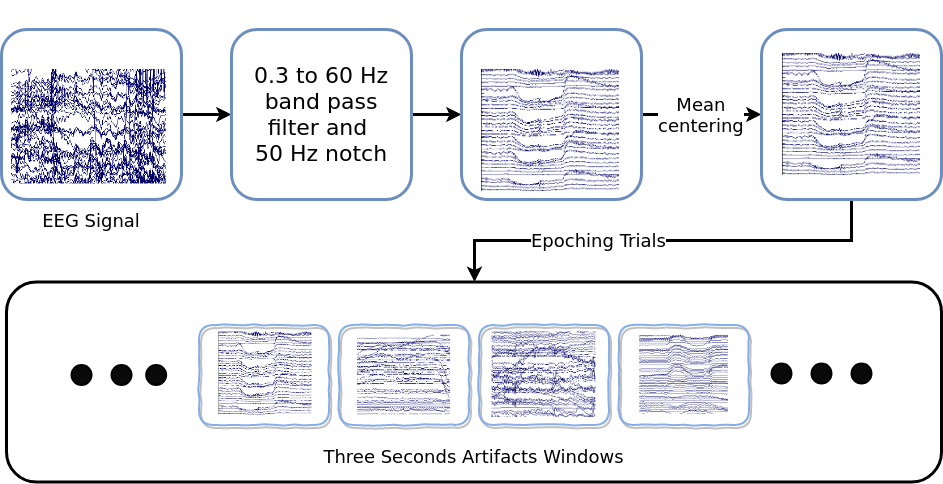}
  \caption{Preprocessing of Electroencephalographical Signal}
  \label{Fig:PreproEEG}
\end{figure*}

Collected EEG data signals were filtered with 0.3 to 60 Hz bandpass filter, and a notch of 50 Hz was employed to overcome the line noise. The resultant EEG signals were mean-centered. Three seconds periods provided for artifact production were extracted from the mean-centered EEG signal, called Epochs. Considering that all the channels are independent, the three seconds EEG signal results in a 128x750 dimension vector given the sampling rate of 250Hz. The illustrative block diagram of the preprocessing of the EEG signal is shown in the Figure~\ref{Fig:PreproEEG}.


\section{Detection of Artifacts in Epochs}
\label{sec:thres}
\begin{figure*}[!ht]
\begin{minipage}[h]{1.0\linewidth}
 \centering
       \includegraphics[scale=0.4]{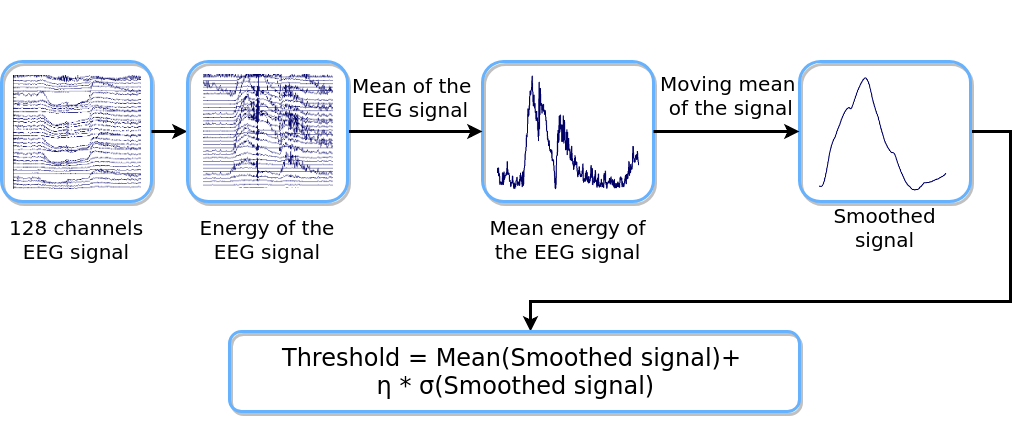}
  \caption{Illustration of artifact detection using 128 channel EEG}
  \label{Fig:FlowChart}
      \end{minipage}
\end{figure*}

Artifacts in the EEG signal influence and exhibit large amplitudes in the electrodes. Utilizing this understanding, we devise a threshold-based procedure to detect the artifacts in the epoched EEG signal. Within a three second time window, a subject voluntarily performs the expected artifact. However, the subject has the liberty to initiate and windup the artifact anytime in three seconds. Moreover, the duration of each artifact might not be the same. To detect the onset and completion of the artifacts in the given three seconds window, we introduced a threshold-based technique. The energy of epoched three seconds EEG signal from all the 128 electrodes was estimated. The mean energy of the 128 electrodes EEG signals was calculated. In order to smoothen the resultant mean energy signal, we applied a moving average filter of length 100 samples (0.4 seconds). This signal proffers a signature of the artifact and will be mentioned hereafter as ``artifact-signal''. Using a threshold on the energy of this smoothed EEG signal, the location of the artifacts in the EEG signal was first localized. The onset and completion of the artifacts in the EEG can be detected by,
\begin{equation*}
    Threshold = mean\left(\text{``artifact-signal''}\right) + \eta * \sigma \left(\text{``artifact-signal''}\right)
\end{equation*}
where $\eta$ is a hyper-parameter and $\sigma$ is the standard deviation of ``artifact-signal''. Empirically, we determined that the hyper-parameter $\eta$ = $-1$ operates adequately for this task. The extent in the amplitude of the smoothened mean energy of the EEG signal passes the threshold determines the onset and completion of the artifacts in the epoched EEG signals. Figure~\ref{Fig:FlowChart} illustrates this process. Once the onset and decay of the artifacts in the EEG have been identified, the type of artifact can be reliably identified using time warping techniques. The time duration of each artifact can differ from trial to trial and person to person. So it is crucial to use time warping. Figure~\ref{Fig:Example} presents the distinction in the time duration of the artifacts for various subjects. 

\section{Classification of Artifacts} 
\label{sec:classArti}
During data acquisition, the subjects were instructed to produce the artifacts in a span of $3$ seconds time window. In the last section, we discussed the technique to detect the onset and completion of the artifacts in the three seconds EEG signal window. In this section, we discuss the methodologies used to classify the discovered artifacts effectively. Since the subject can initiate and complete the artifacts anytime and time taken for artifacts is of varying duration, we adopted time warping techniques to classify the artifacts.

\begin{table}[h]
\centering
\caption{Number of subjects and trials used in the detection and classification of artifacts in 128 electrodes EEG}
\begin{tabular}{c|c|c|c|c|}
\cline{2-5}
                                              & \multicolumn{2}{c|}{\textbf{\begin{tabular}[c]{@{}c@{}}Number of Subjects\end{tabular}}} & \multicolumn{2}{c|}{\textbf{\begin{tabular}[c]{@{}c@{}}Number of Trials\end{tabular}}} \\ \cline{2-5} 
                                              & \textbf{Train}                              & \textbf{Test}                             & \textbf{Train}                             & \textbf{Test}                            \\ \hline
\multicolumn{1}{|c|}{\textbf{Single Session}} &    9                                         &     9                                      &   103                                         &          103                                \\ \hline
\multicolumn{1}{|c|}{\textbf{Inter Sessions}} &   4                                          &          4                                 &   50                                         &        55                                  \\ \hline

\multicolumn{1}{|c|}{\textbf{Inter Subjects}} &   5                                          &          4                                 &       75                                     &      60                                    \\ \hline
\end{tabular}
\label{Tab:TrialsCount}
\end{table}
To validate the robustness of the proposing system, we attempted three separate investigations to identify and classify the artifacts in the EEG signal. The investigations were intra-session, inter-session, and inter-subject classification of the artifacts from the EEG signals and discussed below. The number of subjects and trials utilized in training and testing in various arrangements is compiled in the Table~\ref{Tab:TrialsCount}.

\section{Linear Time Warping algorithm} 
\label{sec:ltw}
Linear Time Warping (LTW) algorithm is a method in sequential pattern recognition where both series are interpolated to the same or fixed length and classified based on any distance metrics like Euclidean distance of the series. The onset and the completion of an artifact in the EEG signal are first identified using the threshold method. Nevertheless, the duration of the artifacts can differ from time to time. We firstly interpolate test and reference series to the max length of the test and the reference series and then examine using Euclidean distance. Before comparing the test and reference series, the variability in the amplitude was normalized. Finally, LTW distance is employed as a score to classify the artifact of a specific type using distance-based $k$-Nearest Neighbour classifiers, as given in Algorithm~\ref{Algo:LTW}.

\begin{algorithm}[!ht]
\setstretch{1.5}
\SetAlgoLined
\KwIn{evaluationTrial, referenceTrials, noOfNeighbours}
\KwResult{predictedLabel}
 predictedLabel := nil\;
 costFromReferenceTrials[] := Inf\;
 \ForAll {referenceTrial in referenceTrials}{
 maxLength := max(len(evaluationTrial), len(referenceTrial))\;
 evaluationTrial := linearInterpolate(evaluationTrial, maxLength)\;
 referenceTrial := linearInterpolate(referenceTrial, maxLength)\;
 costFromReferenceTrials[referenceTrial] := ||evaluationTrial - referenceTrial||\;
 }
 sort(costFromReferenceTrials)\;
 predictedLabel := $argmin$ (costFromReferenceTrials[1 : noOfNeighbours])\;
 \caption{Linear Time Warping algorithm based artifacts classification}
 \label{Algo:LTW}
\end{algorithm}

\subsection*{Intra-Session artifacts classification} 
In this experiment, single-session data of all nine subjects were utilized. From every session, random $50\%$ of the trials were utilized as the reference series, and the remaining are employed as the test series. The outcome of this classification for different values of $k$ is provided in the Table~\ref{Tab:LTWIntra}. We observed that the classification accuracy of eye-blink is highest, followed by jaw movement, head nod, and head turn.  We also observed that head nod and head turn were difficult to distinguish.

\begin{table}[h]
\centering
\caption{Artifacts classification results of Linear Time Warping in Intra-Session}
\begin{tabular}{c|c|c|c|c|c|c|}
\cline{2-7}
                                                    & \textbf{k} & \textbf{\begin{tabular}[c]{@{}c@{}}Jaw \\ Movement\end{tabular}} & \textbf{\begin{tabular}[c]{@{}c@{}}Head \\ Nod\end{tabular}} & \textbf{\begin{tabular}[c]{@{}c@{}}Head \\ Turn\end{tabular}} & \textbf{\begin{tabular}[c]{@{}c@{}}Eye \\ Blink\end{tabular}} & \textbf{\begin{tabular}[c]{@{}c@{}}Model \\ Accuracy\end{tabular}} \\ \hline
\multicolumn{1}{|c|}{\multirow{4}{*}{\textbf{LTW}}} 			     			& \textbf{1}                           & 90.3\%                                 & 74.8\%                                                       & 85.4\%                                                              & 100\%                                                        & \textit{\textbf{87.6\%}}                                           \\ \cline{2-7} 
\multicolumn{1}{|c|}{}                              & \textbf{3}                       & 88.3\%                                 & 70.9\%                                                       & 86.4\%                                                             & 100\%                                                        & \textit{\textbf{86.4\%}}                                           \\ \cline{2-7} 
\multicolumn{1}{|c|}{}                              & \textbf{5}                        & 88.3\%                                 & 70.9\%                                                       & 83.5\%                                                            & 99.0\%                                                        & \textit{\textbf{85.4\%}}                                           \\ \cline{2-7} 
\multicolumn{1}{|c|}{}                              & \textbf{7}                        & 92.2\%                                 & 68.9\%                                                       & 81.6\%                                                            & 99.0\%                                                        & \textit{\textbf{85.4\%}}                                           \\ \hline
\end{tabular}
\label{Tab:LTWIntra}
\end{table}

\subsection*{Inter-Session artifacts classification} In this subsection, the four subjects' data with multiple sessions have been utilized. The arbitrarily taken session data of every subject was utilized as a reference series, and another session data was utilized as a test series. The outcome of this agreement is shown in the Table~\ref{Tab:LTWInterSes}. The records in the table show the individual accuracy of each artifact, obtained by the four-class classification models, accompanying with the total accuracy of the designed classification model.

\begin{table}[h]
\centering
\caption{Artifacts classification results of Linear Time Warping in Inter-Session}
\begin{tabular}{c|c|c|c|c|c|c|}
\cline{2-7}
                                                    & \textbf{k} & \textbf{\begin{tabular}[c]{@{}c@{}}Jaw \\ Movement\end{tabular}} & \textbf{\begin{tabular}[c]{@{}c@{}}Head \\ Nod\end{tabular}} & \textbf{\begin{tabular}[c]{@{}c@{}}Head \\ Turn\end{tabular}} & \textbf{\begin{tabular}[c]{@{}c@{}}Eye \\ Blink\end{tabular}} & \textbf{\begin{tabular}[c]{@{}c@{}}Model \\ Accuracy\end{tabular}} \\ \hline
\multicolumn{1}{|c|}{\multirow{4}{*}{\textbf{LTW}}} 			     & \textbf{1}                      & 78.2\%                                    & 45.4\%                                                       & 81.8\%                                                             & 94.5\%                                                      & \textit{\textbf{75.0\%}}                                           \\ \cline{2-7} 
\multicolumn{1}{|c|}{}                              & \textbf{3}                       & 74.5\%                                   & 52.7\%                                                       & 78.2\%                                                             & 98.2\%                                                      & \textit{\textbf{75.9\%}}                                           \\ \cline{2-7} 
\multicolumn{1}{|c|}{}                              & \textbf{5}                       & 69.1\%                                   & 49.1\%                                                       & 80.0\%                                                             & 96.4\%                                                      & \textit{\textbf{73.6\%}}                                           \\ \cline{2-7} 
\multicolumn{1}{|c|}{}                              & \textbf{7}                       & 70.9\%                                   & 45.5\%                                                       & 80.0\%                                                             & 90.9\%                                                      & \textit{\textbf{71.8\%}}                                           \\ \hline
\end{tabular}
\label{Tab:LTWInterSes}
\end{table}

\subsection*{Inter-Subject artifacts classification} In this subsection, artifact signatures were examined across subjects. From all the nine subjects, four subjects' EEG signal data were arbitrarily picked as the reference series, and the remaining five subjects' EEG signal data were utilized as the test series. The outcome of this agreement is presented in the Table~\ref{Tab:LTWUnseen}. The records in the table show the individual accuracy of each artifact, obtained by the four-class classification models, accompanying with the total accuracy of the designed classification model.

\begin{table}[h]
\centering
\caption{Artifacts classification results of Linear Time Warping in Inter-Subject}
\begin{tabular}{c|c|c|c|c|c|c|}
\cline{2-7}
                                                    & \textbf{k} & \textbf{\begin{tabular}[c]{@{}c@{}}Jaw \\ Movement\end{tabular}} & \textbf{\begin{tabular}[c]{@{}c@{}}Head \\ Nod\end{tabular}} & \textbf{\begin{tabular}[c]{@{}c@{}}Head \\ Turn\end{tabular}} & \textbf{\begin{tabular}[c]{@{}c@{}}Eye \\ Blink\end{tabular}} & \textbf{\begin{tabular}[c]{@{}c@{}}Model \\ Accuracy\end{tabular}} \\ \hline
\multicolumn{1}{|c|}{\multirow{4}{*}{\textbf{LTW}}} 			     & \textbf{1}                            & 86.7\%                          & 11.7\%                                                       & 61.7\%                                                                   & 95.0\%                                                    & \textit{\textbf{63.8\%}}                                           \\ \cline{2-7} 
\multicolumn{1}{|c|}{}                              & \textbf{3}                             & 88.3\%                          & 11.7\%                                                       & 61.7\%                                                                  & 98.3\%                                                    & \textit{\textbf{65.0\%}}                                           \\ \cline{2-7} 
\multicolumn{1}{|c|}{}                              & \textbf{5}                             & 90.0\%                          & 10.0\%                                                       & 61.7\%                                                                 & 98.3\%                                                     & \textit{\textbf{65.0\%}}                                           \\ \cline{2-7} 
\multicolumn{1}{|c|}{}                              & \textbf{7}                            & 88.3\%                           & 1.7\%                                                        & 61.7\%                                                                  & 98.3\%                                                    & \textit{\textbf{62.5\%}}                                           \\ \hline
\end{tabular}
\label{Tab:LTWUnseen}
\end{table}

From the Tables~\ref{Tab:LTWIntra}, \ref{Tab:LTWInterSes}, and \ref{Tab:LTWUnseen}, we observe that there is a graceful degradation in performance.  Nevertheless, the performance is much better than chance.  This suggests that artifact signatures are indeed present in the EEG signal.

\section{Dynamic Time Warping algorithm}
\label{sec:dtw}
In contrast to LTW, Dynamic Time Warping (DTW) algorithm attempts to match a non-linear warp between the test and the reference series \citep{sakoe1978dynamic}. This warping between two time series can be utilized to ascertain the similarities and differences between the two time series by computing the DTW distance. The DTW distance is employed as a score to classify the artifact of a specific type using distance-based $k$-Nearest Neighbour classifiers, as given in the Algorithm~\ref{Algo:DTW}. 

\begin{algorithm}[!ht]
\setstretch{1.5}
\SetAlgoLined
\KwIn{evaluationTrial, referenceTrials, noOfNeighbours}
\KwResult{predictedLabel}
 leastCost := Inf\;
 predictedLabel := nil\;
 costFromReferenceTrials[] := Inf\;
 refLength := length(referenceTrial)\;
 evalLength := length(evaluationTrial)\;
 \ForAll {referenceTrial in referenceTrials}{
 \For{iter1 := 1 to refLength}{DTW[iter1, 0] := Inf\;}
 \For{iter2 := 1 to evalLength}{DTW[0, iter2] := Inf\;}
 DTW[0, 0] := 0\;
 \For{iter1 := 1 to refLength}{
 \For{iter2 := 1 to evalLength}{
 cost := d(referenceTrial[iter1], evaluationTrial[iter2])\;
 DTW[iter1, iter2] := cost + $min$(DTW[iter1 - 1, iter2  ], 
 \\\hspace*{5em}DTW[iter1, iter2 - 1], DTW[iter1 - 1, iter2 - 1])\;
 
 }
 }
 costFromReferenceTrials[referenceTrial] := DTW[refLength, evalLength]\;
 }
 sort(costFromReferenceTrials)\;
 predictedLabel := $argmin$ (costFromReferenceTrials[1 : noOfNeighbours])\;
 \caption{Dynamic Time Warping algorithm based artifacts classification}
 \label{Algo:DTW}
\end{algorithm}

The following sections deal with variants of the DTW algorithm and their vocation to classify the artifacts in the EEG signals. The data split of artifacts in EEG signals for classifications is identical to that of experiments on LTW, as mentioned in Section~\ref{sec:ltw}.

%

\subsection{Vanilla Dynamic Time Warping algorithm}
\label{SubSec:SimDTW}
A simple Dynamic Time Warping computes the DTW distance between two given series. The following equation determines the DTW distance of two signals.

\begin{equation}
\label{Eqn:DTW}
r_{i,j} = dist \left( \bar{x}_i, \bar{y}_j \right) + min \left( r_{i-1,j-1}, r_{i-1,j},r_{i,j-1} \right)
\end{equation}
where $\bar{x}_i$ and $\bar{y}_j$ are features of the two time series signals compared, $dist\left( \bar{x}_i, \bar{y}_j \right)$ is the Euclidean distance between $\bar{x}_i$ and $\bar{y}_j$. $r_{0,0}$ is initialized with 0 and remaining $r_{i,j}$ are initialized with infinity. The signals are normalized before comparing. Finally, DTW distance is employed as a score to classify or detect the artifact of a specific type using distance-based $k$-Nearest Neighbour classifier. The classification accuracies of intra-session, inter-session, and inter-subject are shown in the Tables~\ref{Tab:SimDTWIntra}, \ref{Tab:SimDTWInter}, and \ref{Tab:SimDTWUnseen} respectively. The records in the tables show the individual accuracy of each artifact, obtained by the four-class classification models, accompanying with the total accuracy of the designed classification model.

\begin{table}[h]
\centering
\caption{Artifacts classification results of vanilla Dynamic Time Warping in Intra-Session}
\begin{tabular}{c|c|c|c|c|c|c|}
\cline{2-7}
                                                    & \textbf{k} & \textbf{\begin{tabular}[c]{@{}c@{}}Jaw \\ Movement\end{tabular}} & \textbf{\begin{tabular}[c]{@{}c@{}}Head \\ Nod\end{tabular}} & \textbf{\begin{tabular}[c]{@{}c@{}}Head \\ Turn\end{tabular}} & \textbf{\begin{tabular}[c]{@{}c@{}}Eye \\ Blink\end{tabular}} & \textbf{\begin{tabular}[c]{@{}c@{}}Model \\ Accuracy\end{tabular}} \\ \hline
\multicolumn{1}{|c|}{\multirow{4}{*}{\textbf{Vanilla DTW}}} 			     & \textbf{1}                         & 85.9\%                                  & 91.0\%                                                       & 92.3\%                                                              & 91.3\%                                                    & \textit{\textbf{87.6\%}}                                           \\ \cline{2-7} 
\multicolumn{1}{|c|}{}                              & \textbf{3}                           & 83.3\%                                 & 82.1\%                                                       & 87.2\%                                                             & 91.3\%                                                    & \textit{\textbf{87.2\%}}                                           \\ \cline{2-7} 
\multicolumn{1}{|c|}{}                              & \textbf{5}                           & 85.9\%                                 & 71.8\%                                                       & 87.2\%                                                             & 91.3\%                                                    & \textit{\textbf{85.2\%}}                                           \\ \cline{2-7} 
\multicolumn{1}{|c|}{}                              & \textbf{7}                           & 84.6\%                                 & 65.4\%                                                       & 80.8\%                                                             & 91.3\%                                                    & \textit{\textbf{81.4\%}}                                           \\ \hline
\end{tabular}
\label{Tab:SimDTWIntra}
\end{table}

\begin{table}[h]
\centering
\caption{Artifacts classification results of vanilla Dynamic Time Warping in Inter-Session}
\begin{tabular}{c|c|c|c|c|c|c|}
\cline{2-7}
                                                    & \textbf{k} & \textbf{\begin{tabular}[c]{@{}c@{}}Jaw \\ Movement\end{tabular}} & \textbf{\begin{tabular}[c]{@{}c@{}}Head \\ Nod\end{tabular}} & \textbf{\begin{tabular}[c]{@{}c@{}}Head \\ Turn\end{tabular}} & \textbf{\begin{tabular}[c]{@{}c@{}}Eye \\ Blink\end{tabular}} & \textbf{\begin{tabular}[c]{@{}c@{}}Model \\ Accuracy\end{tabular}} \\ \hline
\multicolumn{1}{|c|}{\multirow{4}{*}{\textbf{Vanilla DTW}}} & \textbf{1} 				                               & 49.1\%                                     & 47.3\%                                                       & 48.3\%                                                                & 91.3\%                                              & \textit{\textbf{59.0\%}}                                           \\ \cline{2-7} 
\multicolumn{1}{|c|}{}                              & \textbf{3}                          & 47.3\%                                     & 40.0\%                                                       & 51.7\%                                                                & 91.3\%                                              & \textit{\textbf{57.6\%}}                                           \\ \cline{2-7} 
\multicolumn{1}{|c|}{}                              & \textbf{5}                          & 40.0\%                                     & 40.0\%                                                       & 53.3\%                                                                & 91.3\%                                              & \textit{\textbf{56.2\%}}                                           \\ \cline{2-7} 
\multicolumn{1}{|c|}{}                              & \textbf{7}                          & 40.9\%                                     & 40.0\%                                                       & 51.6\%                                                                & 91.3\%                                              & \textit{\textbf{55.9\%}}                                           \\ \hline
\end{tabular}
\label{Tab:SimDTWInter}
\end{table}

\begin{table}[h]
\centering
\caption{Artifacts classification results of vanilla Dynamic Time Warping in Inter-Subject}
\begin{tabular}{c|c|c|c|c|c|c|}
\cline{2-7}
                                                    & \textbf{k} & \textbf{\begin{tabular}[c]{@{}c@{}}Jaw \\ Movement\end{tabular}} & \textbf{\begin{tabular}[c]{@{}c@{}}Head \\ Nod\end{tabular}} & \textbf{\begin{tabular}[c]{@{}c@{}}Head \\ Turn\end{tabular}} & \textbf{\begin{tabular}[c]{@{}c@{}}Eye \\ Blink\end{tabular}} & \textbf{\begin{tabular}[c]{@{}c@{}}Model \\ Accuracy\end{tabular}} \\ \hline
\multicolumn{1}{|c|}{\multirow{4}{*}{\textbf{Vanilla DTW}}} & \textbf{1} 				                                & 67.7\%                                & 10.0\%                                                       & 48.3\%                                                      & 91.3\%                                                            & \textit{\textbf{54.3\%}}                                           \\ \cline{2-7} 
\multicolumn{1}{|c|}{}                              & \textbf{3}                         & 66.3\%                                  & 6.7\%                                                       & 51.7\%                                                        & 91.3\%                                                          & \textit{\textbf{54.0\%}}                                           \\ \cline{2-7} 
\multicolumn{1}{|c|}{}                              & \textbf{5}                         & 70.0\%                                  & 5.3\%                                                       & 47.7\%                                                        & 91.3\%                                                          & \textit{\textbf{53.6\%}}                                           \\ \cline{2-7} 
\multicolumn{1}{|c|}{}                              & \textbf{7}                          & 70.1\%                                 & 4.7\%                                                        & 51.7\%                                                       & 91.3\%                                                           & \textit{\textbf{54.4\%}}                                           \\ \hline
\end{tabular}
\label{Tab:SimDTWUnseen}
\end{table}

\subsection{Normalized Dynamic Time Warping algorithm}
The normalized Dynamic Time Warping (DTW) is one of the variants of the DTW algorithm. The DTW matrix computation is identical, as mentioned in subsection~\ref{SubSec:SimDTW}. However, the final DTW distances are normalized by the length of the warping path. The normalized DTW distance is employed as a score to classify the artifact of a specific type using distance-based $k$-Nearest Neighbour classifier. The classification accuracies of intra-session, inter-session, and inter-subject are shown in the Tables~\ref{Tab:NorDTWIntra}, \ref{Tab:NorDTWInter}, and \ref{Tab:NorDTWUnseen} respectively. The records in the tables show the individual accuracy of each artifact, obtained by the four-class classification models, accompanying with the total accuracy of the designed classification model.

\begin{table}[h]
\centering
\caption{Artifacts classification results of normalized Dynamic Time Warping in Intra-Session}
\begin{tabular}{c|c|c|c|c|c|c|}
\cline{2-7}
                                                    & \textbf{k} & \textbf{\begin{tabular}[c]{@{}c@{}}Jaw \\ Movement\end{tabular}} & \textbf{\begin{tabular}[c]{@{}c@{}}Head \\ Nod\end{tabular}} & \textbf{\begin{tabular}[c]{@{}c@{}}Head \\ Turn\end{tabular}} & \textbf{\begin{tabular}[c]{@{}c@{}}Eye \\ Blink\end{tabular}} & \textbf{\begin{tabular}[c]{@{}c@{}}Model \\ Accuracy\end{tabular}} \\ \hline
\multicolumn{1}{|c|}{\multirow{4}{*}{\textbf{Normalized DTW}}} & \textbf{1} 				                             & 87.4\%                        & 83.5\%                                                       & 92.2\%                                                                    & 100\%                                                         & \textit{\textbf{90.8\%}}                                           \\ \cline{2-7} 
\multicolumn{1}{|c|}{}                              & \textbf{3}                      & 84.5\%                        & 80.6\%                                                       & 88.3\%                                                                      & 100\%                                                        & \textit{\textbf{88.3\%}}                                           \\ \cline{2-7} 
\multicolumn{1}{|c|}{}                              & \textbf{5}                      & 88.3\%                        & 81.6\%                                                       & 86.4\%                                                                      & 100\%                                                        & \textit{\textbf{89.1\%}}                                           \\ \cline{2-7} 
\multicolumn{1}{|c|}{}                              & \textbf{7}                       & 89.3\%                        & 78.6\%                                                       & 85.4\%                                                                     & 99.0\%                                                        & \textit{\textbf{88.1\%}}                                           \\ \hline
\end{tabular}
\label{Tab:NorDTWIntra}
\end{table}

\begin{table}[h]
\centering
\caption{Artifacts classification results of normalized Dynamic Time Warping in Inter-Session}
\begin{tabular}{c|c|c|c|c|c|c|}
\cline{2-7}
                                                    & \textbf{k} & \textbf{\begin{tabular}[c]{@{}c@{}}Jaw \\ Movement\end{tabular}} & \textbf{\begin{tabular}[c]{@{}c@{}}Head \\ Nod\end{tabular}} & \textbf{\begin{tabular}[c]{@{}c@{}}Head \\ Turn\end{tabular}} & \textbf{\begin{tabular}[c]{@{}c@{}}Eye \\ Blink\end{tabular}} & \textbf{\begin{tabular}[c]{@{}c@{}}Model \\ Accuracy\end{tabular}} \\ \hline
\multicolumn{1}{|c|}{\multirow{4}{*}{\textbf{Normalized DTW}}} & \textbf{1} 				                                       & 47.3\%                     & 49.1\%                                                       & 67.3\%                                                                & 98.1\%                                                     & \textit{\textbf{65.5\%}}                                           \\ \cline{2-7} 
\multicolumn{1}{|c|}{}                              & \textbf{3}                                 & 40.0\%                     & 47.3\%                                                       & 67.3\%                                                                 & 100\%                                                     & \textit{\textbf{63.6\%}}                                           \\ \cline{2-7} 
\multicolumn{1}{|c|}{}                              & \textbf{5}                                 & 38.2\%                     & 40.0\%                                                       & 70.9\%                                                                 & 100\%                                                     & \textit{\textbf{62.3\%}}                                           \\ \cline{2-7} 
\multicolumn{1}{|c|}{}                              & \textbf{7}                                 & 43.6\%                     & 40.0\%                                                       & 70.9\%                                                                 & 100\%                                                     & \textit{\textbf{63.6\%}}                                           \\ \hline
\end{tabular}
\label{Tab:NorDTWInter}
\end{table}

\begin{table}[]
\centering
\caption{Artifacts classification results of normalized Dynamic Time Warping in Inter-Subject}
\begin{tabular}{c|c|c|c|c|c|c|}
\cline{2-7}
                                                    & \textbf{k} & \textbf{\begin{tabular}[c]{@{}c@{}}Jaw \\ Movement\end{tabular}} & \textbf{\begin{tabular}[c]{@{}c@{}}Head \\ Nod\end{tabular}} & \textbf{\begin{tabular}[c]{@{}c@{}}Head \\ Turn\end{tabular}} & \textbf{\begin{tabular}[c]{@{}c@{}}Eye \\ Blink\end{tabular}} & \textbf{\begin{tabular}[c]{@{}c@{}}Model \\ Accuracy\end{tabular}} \\ \hline
\multicolumn{1}{|c|}{\multirow{4}{*}{\textbf{Normalized DTW}}} & \textbf{1} 				                            & 80.0\%                  & 5.0\%                                                       & 48.3\%                                                                              & 93.3\%                                                     & \textit{\textbf{56.7\%}}                                           \\ \cline{2-7} 
\multicolumn{1}{|c|}{}                              & \textbf{3}                     & 75.0\%                  & 6.7\%                                                       & 51.7\%                                                                                & 93.3\%                                                     & \textit{\textbf{56.7\%}}                                           \\ \cline{2-7} 
\multicolumn{1}{|c|}{}                              & \textbf{5}                     & 76.7\%                  & 8.3\%                                                       & 53.3\%                                                                                & 93.3\%                                                     & \textit{\textbf{56.1\%}}                                           \\ \cline{2-7} 
\multicolumn{1}{|c|}{}                              & \textbf{7}                     & 76.6\%                  & 1.6\%                                                       & 51.6\%                                                                                & 93.3\%                                                     & \textit{\textbf{55.8\%}}                                           \\ \hline
\end{tabular}
\label{Tab:NorDTWUnseen}
\end{table}

\subsection{Time Synchronized Dynamic Time Warping algorithm}

The time synchronized Dynamic Time Warping (DTW) computes the DTW distance between two given series. The following equation determines the DTW distance of two signals.
\begin{equation}
\label{Eqn:DTW}
r_{i,j} = dist \left( \bar{x}_i, \bar{y}_j \right) + min \left( r_{i-1,j-1}, r_{i-1,j} \right)
\end{equation}
where $\bar{x}_i$ and $\bar{y}_j$ are features of the two time series signals compared, $dist\left( \bar{x}_i, \bar{y}_j \right)$ is the Euclidean distance between $\bar{x}_i$ and $\bar{y}_j$. The initialization of the $r$ matrix is as same as mentioned in Subsection~\ref{SubSec:SimDTW}. Here, the warping is restricted to vertical and diagonal. The signals are normalized to zero mean and unit variance before comparing. The estimated DTW distance is employed as a score to classify or detect the artifact of a specific type using distance-based $k$-Nearest Neighbour classifier. The classification accuracies of intra-session, inter-session, and inter-subject are shown in the Tables~\ref{Tab:TimeDTWIntra}, \ref{Tab:TimeDTWInter}, and \ref{Tab:TimeDTWUnseen} respectively. The records in the tables show the individual accuracy of each artifact, obtained by the four-class classification models, accompanying with the total accuracy of the designed classification model.

\begin{table}[h]
\centering
\caption{Artifacts classification results of time synchronized Dynamic Time Warping in Intra-Session}
\begin{tabular}{c|c|c|c|c|c|c|}
\cline{2-7}
                                                    & \textbf{k} & \textbf{\begin{tabular}[c]{@{}c@{}}Jaw \\ Movement\end{tabular}} & \textbf{\begin{tabular}[c]{@{}c@{}}Head \\ Nod\end{tabular}} & \textbf{\begin{tabular}[c]{@{}c@{}}Head \\ Turn\end{tabular}} & \textbf{\begin{tabular}[c]{@{}c@{}}Eye \\ Blink\end{tabular}} & \textbf{\begin{tabular}[c]{@{}c@{}}Model \\ Accuracy\end{tabular}} \\ \hline
\multicolumn{1}{|c|}{\multirow{4}{*}{\textbf{Time Synchronized DTW}}} & \textbf{1} 				                              & 85.9\%                             & 83.5\%                                                       & 92.2\%                                                                             & 98.2\%                                          & \textit{\textbf{89.9\%}}                                           \\ \cline{2-7} 
\multicolumn{1}{|c|}{}                              & \textbf{3}                         & 83.3\%                             & 80.6\%                                                       & 87.2\%                                                                             & 98.2\%                                          & \textit{\textbf{86.4\%}}                                           \\ \cline{2-7} 
\multicolumn{1}{|c|}{}                              & \textbf{5}                         & 85.9\%                             & 71.8\%                                                       & 86.4\%                                                                             & 98.2\%                                          & \textit{\textbf{85.5\%}}                                           \\ \cline{2-7} 
\multicolumn{1}{|c|}{}                              & \textbf{7}                         & 84.6\%                             & 65.4\%                                                       & 80.8\%                                                                             & 98.2\%                                          & \textit{\textbf{82.3\%}}                                           \\ \hline
\end{tabular}
\label{Tab:TimeDTWIntra}
\end{table}

\begin{table}[h]
\centering
\caption{Artifacts classification results of time synchronized Dynamic Time Warping in Inter-Session}
\begin{tabular}{c|c|c|c|c|c|c|}
\cline{2-7}
                                                    & \textbf{k} & \textbf{\begin{tabular}[c]{@{}c@{}}Jaw \\ Movement\end{tabular}} & \textbf{\begin{tabular}[c]{@{}c@{}}Head \\ Nod\end{tabular}} & \textbf{\begin{tabular}[c]{@{}c@{}}Head \\ Turn\end{tabular}} & \textbf{\begin{tabular}[c]{@{}c@{}}Eye \\ Blink\end{tabular}} & \textbf{\begin{tabular}[c]{@{}c@{}}Model \\ Accuracy\end{tabular}} \\ \hline
\multicolumn{1}{|c|}{\multirow{4}{*}{\textbf{Time Synchronized DTW}}} & \textbf{1} 				                                & 47.3\%                                   & 47.3\%                                                       & 48.3\%                                                             & 94.5\%                                                  & \textit{\textbf{59.6\%}}                                           \\ \cline{2-7} 
\multicolumn{1}{|c|}{}                              & \textbf{3}                           & 40.0\%                                   & 40.0\%                                                       & 51.7\%                                                            & 94.5\%                                                   & \textit{\textbf{56.5\%}}                                           \\ \cline{2-7} 
\multicolumn{1}{|c|}{}                              & \textbf{5}                           & 38.2\%                                   & 40.0\%                                                       & 53.3\%                                                            & 94.5\%                                                   & \textit{\textbf{56.5\%}}                                           \\ \cline{2-7} 
\multicolumn{1}{|c|}{}                              & \textbf{7}                           & 40.9\%                                   & 40.0\%                                                       & 51.6\%                                                            & 90.9\%                                                   & \textit{\textbf{55.8\%}}                                           \\ \hline
\end{tabular}
\label{Tab:TimeDTWInter}
\end{table}

\begin{table}[]
\centering
\caption{Artifacts classification results of time synchronized Dynamic Time Warping in Inter-Subject}
\begin{tabular}{c|c|c|c|c|c|c|}
\cline{2-7}
                                                    & \textbf{k} & \textbf{\begin{tabular}[c]{@{}c@{}}Jaw \\ Movement\end{tabular}} & \textbf{\begin{tabular}[c]{@{}c@{}}Head \\ Nod\end{tabular}} & \textbf{\begin{tabular}[c]{@{}c@{}}Head \\ Turn\end{tabular}} & \textbf{\begin{tabular}[c]{@{}c@{}}Eye \\ Blink\end{tabular}} & \textbf{\begin{tabular}[c]{@{}c@{}}Model \\ Accuracy\end{tabular}} \\ \hline
\multicolumn{1}{|c|}{\multirow{4}{*}{\textbf{Time Synchronized DTW}}} & \textbf{1} 				                                     & 67.7\%                              & 5.7\%                                                       & 48.3\%                                                              & 91.0\%                                                 & \textit{\textbf{53.2\%}}                                           \\ \cline{2-7} 
\multicolumn{1}{|c|}{}                              & \textbf{3}                               & 66.8\%                              & 6.7\%                                                       & 50.3\%                                                               & 91.0\%                                                 & \textit{\textbf{53.7\%}}                                           \\ \cline{2-7} 
\multicolumn{1}{|c|}{}                              & \textbf{5}                               & 70.0\%                              & 8.3\%                                                       & 48.4\%                                                               & 91.0\%                                                 & \textit{\textbf{54.4\%}}                                           \\ \cline{2-7} 
\multicolumn{1}{|c|}{}                              & \textbf{7}                                & 70.6\%                             & 2.3\%                                                        & 51.4\%                                                              & 91.0\%                                                  & \textit{\textbf{53.8\%}}                                           \\ \hline
\end{tabular}
\label{Tab:TimeDTWUnseen}
\end{table}

From the Tables~\ref{Tab:SimDTWIntra}, \ref{Tab:SimDTWInter},  \ref{Tab:SimDTWUnseen}, \ref{Tab:NorDTWIntra}, \ref{Tab:NorDTWInter}, \ref{Tab:NorDTWUnseen}, \ref{Tab:TimeDTWIntra}, \ref{Tab:TimeDTWInter}, and \ref{Tab:TimeDTWUnseen}, it is evident that the normalized DTW outperformed the task in hand. This phenomenon is an outcome of the fact that the normalized DTW takes the length of the sequences into consideration and whereas other variants of the DTW are indifferent to the length. For example, if two arbitrary artifacts are shorter in duration, the DTW distance between those would be smaller, irrespective of the class, which is a downside to the kNN classifier.

From the tables~\ref{Tab:NorDTWIntra}, \ref{Tab:NorDTWInter}, and \ref{Tab:NorDTWUnseen}, it can be remarked that the proposed method classifies the artifacts with high accuracy of $90.8\%$ while all sessions are utilized for both training and testing. The accuracy drops to $75.9\%$ while tested across sessions and further decreases to $63.8\%$ during the test across subjects. It is crucial to note that only the ``Head Nod'' class is classified with reduced accuracy. The principal cause for this phenomenon is that both head turn and head nod affect identical collection of electrodes (in Figure~\ref{Fig:Example}), and consequently, their signatures are not classifiable across sessions and subjects. Furthermore, LTW provides more reliable efficiency in the inter-session and inter-subject environment. This result emphasizes that the time warping in the EEG signal artifacts is more linear rather than non-linear.

\section{Detection on Continous EEG data} 
\label{sec:det}
In the earlier sections, we used the information about $3$ seconds of artifacts window, where the subject was informed to make the artifacts, were extracted as epochs, and threshold-based detection for the onset and decay of the artifacts were employed. However, in the real world setup, prior information about the occurrence of artifacts would not be accessible. So, in this section, we attempt to tackle a perplexing problem of detection and classification of artifacts in the continuous EEG signal.

In this section, the entire acquired EEG data was used, and the onset and completion of all the artifacts were identified using the threshold technique, as discussed in Section~\ref{sec:thres}. The detected artifacts were considered to be true if the period between detected onset and completion of the artifacts overlie 60\% or more with the $3$ seconds of artifacts ground truth window. The hyper-parameter $\eta$ was tuned, and it was found that a value of $0.8$ and $0.9$ operates most reliably in the setup, and this has been given in the table~\ref{Tab:F1Score}. In the rest of the subsection, $\eta$ = $0.8$ was accepted.

      \begin{table}[]
\centering
\caption{F1 scores in detecting artifacts window from all EEG subjects}
\begin{tabular}{|c|c|c|c|c|c|c|c|c|}
\hline
\textbf{$\eta$}             & -0.1  & 0       & 0.1 & 0.2   & 0.3    & 0.4     & 0.5 & 0.6                  \\ \hline
\textbf{F1} & 79.79\% & 81.23\% & 84.41\% & 87.10\% & 90.05\% & 91.60\%  & 92.75\% & 93.88\%     \\ \hline
\hline
\textbf{$\eta$}                    & 0.7    & \textbf{0.8}     & \textbf{0.9}     & 1.0       & 1.1    & 1.2     & 1.3 & 1.4     \\ \hline
\textbf{F1}  & 94.44\%  & \textbf{94.94}\% & \textbf{94.94\%} & 92.74\%   & 91.89\% & 89.34\% & 86.96\% & 83.19\% \\ \hline
\end{tabular}
\label{Tab:F1Score}
\end{table}

\begin{table}[]
\centering
\caption{Artifacts classification results of Linear Time Warping and normalized Dynamic Time Warping in Intra-Session after detection}
\begin{tabular}{c|c|c|c|c|c|c|}
\cline{2-7}
                                                    & \textbf{k} & \textbf{\begin{tabular}[c]{@{}c@{}}Jaw \\ Movement\end{tabular}} & \textbf{\begin{tabular}[c]{@{}c@{}}Head \\ Nod\end{tabular}} & \textbf{\begin{tabular}[c]{@{}c@{}}Head \\ Turn\end{tabular}} & \textbf{\begin{tabular}[c]{@{}c@{}}Eye \\ Blink\end{tabular}} & \textbf{\begin{tabular}[c]{@{}c@{}}Model \\ Accuracy\end{tabular}} \\ \hline
\multicolumn{1}{|c|}{\multirow{4}{*}{\textbf{LTW}}} & \textbf{1} 				                         & 74.8\%                              & 52.4\%                                                       & 79.6\%                                                                            & 75.7\%                                              & \textit{\textbf{70.6\%}}                                           \\ \cline{2-7} 
\multicolumn{1}{|c|}{}                              & \textbf{3}                     & 71.8\%                              & 36.9\%                                                       & 79.6\%                                                                          & 78.6\%                                               & \textit{\textbf{66.8\%}}                                           \\ \cline{2-7} 
\multicolumn{1}{|c|}{}                              & \textbf{5}                     & 71.8\%                              & 37.9\%                                                       & 83.5\%                                                                          & 78.6\%                                               & \textit{\textbf{68.0\%}}                                           \\ \cline{2-7} 
\multicolumn{1}{|c|}{}                              & \textbf{7}                     & 72.8\%                              & 34.9\%                                                       & 83.5\%                                                                          & 78.6\%                                               & \textit{\textbf{67.5\%}}                                           \\ \hline
\multicolumn{1}{|c|}{\multirow{4}{*}{\textbf{Normalized DTW}}} & \textbf{1} 				                                      & 88.4\%                            & 92.2\%                                                       & 91.3\%                                                                              & 91.3\%                                 & \textit{\textbf{90.8\%}}                                           \\ \cline{2-7} 
\multicolumn{1}{|c|}{}                              & \textbf{3}                                 & 87.4\%                            & 90.3\%                                                       & 93.2\%                                                                             & 90.3\%                                  & \textit{\textbf{90.3\%}}                                           \\ \cline{2-7} 
\multicolumn{1}{|c|}{}                              & \textbf{5}                                  & 84.5\%                           & 90.3\%                                                       & 93.2\%                                                                             & 90.3\%                                   & \textit{\textbf{89.3\%}}                                           \\ \cline{2-7} 
\multicolumn{1}{|c|}{}                              & \textbf{7}                                 & 79.6\%                            & 88.4\%                                                       & 92.2\%                                                                             & 90.3\%                                  & \textit{\textbf{87.6\%}}                                           \\ \hline
\end{tabular}
\label{Tab:DetectionClassificationSeen}
\end{table}

\begin{table}[]
\centering
\caption{Artifacts classification results of Linear Time Warping and normalized Dynamic Time Warping in Inter-Session after detection}
\begin{tabular}{c|c|c|c|c|c|c|}
\cline{2-7}
                                                    & \textbf{k} & \textbf{\begin{tabular}[c]{@{}c@{}}Jaw \\ Movement\end{tabular}} & \textbf{\begin{tabular}[c]{@{}c@{}}Head \\ Nod\end{tabular}} & \textbf{\begin{tabular}[c]{@{}c@{}}Head \\ Turn\end{tabular}} & \textbf{\begin{tabular}[c]{@{}c@{}}Eye \\ Blink\end{tabular}} & \textbf{\begin{tabular}[c]{@{}c@{}}Model \\ Accuracy\end{tabular}} \\ \hline
\multicolumn{1}{|c|}{\multirow{4}{*}{\textbf{LTW}}} & \textbf{1} 				                              & 63.6\%                                  & 18.2\%                                                       & 78.2\%                                                                        & 89.1\%                                        & \textit{\textbf{62.3\%}}                                           \\ \cline{2-7} 
\multicolumn{1}{|c|}{}                              & \textbf{3}                           & 74.6\%                                  & 23.6\%                                                       & 80.0\%                                                                      & 89.1\%                                         & \textit{\textbf{66.8\%}}                                           \\ \cline{2-7} 
\multicolumn{1}{|c|}{}                              & \textbf{5}                           & 72.7\%                                  & 16.4\%                                                       & 74.6\%                                                                      & 89.1\%                                         & \textit{\textbf{68.2\%}}                                           \\ \cline{2-7} 
\multicolumn{1}{|c|}{}                              & \textbf{7}                           & 72.7\%                                  & 14.6\%                                                       & 74.6\%                                                                      & 89.1\%                                         & \textit{\textbf{68.2\%}}                                           \\ \hline
\multicolumn{1}{|c|}{\multirow{4}{*}{\textbf{Normalized DTW}}} & \textbf{1} 				                                             & 36.4\%                                 & 38.2\%                                                       & 74.6\%                                                                   & 89.1\%                                & \textit{\textbf{59.6\%}}                                           \\ \cline{2-7} 
\multicolumn{1}{|c|}{}                              & \textbf{3}                                       & 32.7\%                                  & 32.7\%                                                       & 76.4\%                                                                   & 90.9\%                                & \textit{\textbf{58.2\%}}                                           \\ \cline{2-7} 
\multicolumn{1}{|c|}{}                              & \textbf{5}                                        & 29.1\%                                 & 20.0\%                                                       & 78.2\%                                                                   & 90.9\%                                 & \textit{\textbf{55.0\%}}                                           \\ \cline{2-7} 
\multicolumn{1}{|c|}{}                              & \textbf{7}                                       & 29.1\%                                  & 14.6\%                                                       & 78.2\%                                                                   & 90.9\%                                & \textit{\textbf{53.2\%}}                                           \\ \hline
\end{tabular}
\label{Tab:DetectionClassificationSession}
\end{table}

\begin{table}[]
\centering
\caption{Artifacts classification results of Linear Time Warping and normalized Dynamic Time Warping in Inter-Subject after detection}
\begin{tabular}{c|c|c|c|c|c|c|}
\cline{2-7}
                                                    & \textbf{k} & \textbf{\begin{tabular}[c]{@{}c@{}}Jaw \\ Movement\end{tabular}} & \textbf{\begin{tabular}[c]{@{}c@{}}Head \\ Nod\end{tabular}} & \textbf{\begin{tabular}[c]{@{}c@{}}Head \\ Turn\end{tabular}} & \textbf{\begin{tabular}[c]{@{}c@{}}Eye \\ Blink\end{tabular}} & \textbf{\begin{tabular}[c]{@{}c@{}}Model \\ Accuracy\end{tabular}} \\ \hline
\multicolumn{1}{|c|}{\multirow{4}{*}{\textbf{LTW}}} & \textbf{1} &87.7\% &12.3\%&40.0\%&87.7\%&     \textit{\textbf{56.9\%}}                                   \\ \cline{2-7} 
\multicolumn{1}{|c|}{}                              & \textbf{3}&76.9\% & 20.0\% &41.5\%&89.2\%&     \textit{\textbf{56.9\%}}                                   \\ \cline{2-7} 
\multicolumn{1}{|c|}{}                              & \textbf{5}&69.2\% &15.4\%&53.9\%&87.7\%&     \textit{\textbf{56.5\%}}                                       \\ \cline{2-7} 
\multicolumn{1}{|c|}{}                              & \textbf{7} &67.7\% &4.6\%&53.9\%&89.2\%&     \textit{\textbf{53.9\%}}                                       \\ \hline
\multicolumn{1}{|c|}{\multirow{4}{*}{\textbf{Normalized DTW}}} & \textbf{1} 				                                  & 81.5\%                           & 9.2\%                                                       & 40.0\%                                                                     & 86.2\%                                               & \textit{\textbf{54.2\%}}                                           \\ \cline{2-7} 
\multicolumn{1}{|c|}{}                              & \textbf{3}                            & 81.5\%                            & 7.7\%                                                       & 47.7\%                                                                     & 86.2\%                                               & \textit{\textbf{55.8\%}}                                           \\ \cline{2-7} 
\multicolumn{1}{|c|}{}                              & \textbf{5}                              & 81.5\%                          & 9.2\%                                                        & 49.2\%                                                                    & 87.7\%                                                 & \textit{\textbf{56.9\%}}                                           \\ \cline{2-7} 
\multicolumn{1}{|c|}{}                              & \textbf{7}                               & 80.0\%                         & 10.8\%                                                        & 49.2\%                                                                  & 87.7\%                                                  & \textit{\textbf{56.9\%}}                                           \\ \hline
\end{tabular}
\label{Tab:DetectionClassification}
\end{table}

From the last section, it was evident that the normalized Dynamic Time Warping (DTW) works well among it's other variants. So, we attempted normalized DTW along with Linear Time Warping (LTW) for comparison in this section. The outcomes are presented for {\it Intra-Session}, {\it Inter-Session} and {\it Inter-Subject} conditions in Tables~\ref{Tab:DetectionClassificationSeen}, \ref{Tab:DetectionClassificationSession}, and \ref{Tab:DetectionClassification}, respectively. From the Table~\ref{Tab:DetectionClassificationSeen}, it is obvious that the proposed model works well for seen subjects with efficiency around 80\%. The performance of {\it inter-sessions} and {\it inter-subjects} are presented in the Tables~\ref{Tab:DetectionClassificationSession} and \ref{Tab:DetectionClassification}. From the outcomes, it is perceived that despite the performance depravity in {\it inter-sessions} and {\it inter-subjects}, the pattern of the EEG artifacts is consistent. Furthermore, given that the efficiency for {\it Intra-subject}, the artifacts can be productively employed in BCIs.

The performance of LTW was reasonably consistent with varying nearest neighbors in the kNN classifier. This phenomenon shows that the signals were well distributed in LTW metric space. 
Vanilla DTW performs adequately in the {\it Intra-Session} study. However, LTW outperforms the vanilla DTW in {\it Inter-Session} and {\it Inter-Subject} analysis. 
Normalized DTW yields remarkable results in {\it Intra-Session} setup. Despite that, LTW outperforms the normalized DTW. 
Time synchronized DTW yields comparable results in varying neighbors in the classifier. So, the metric space of time synchronized DTW is robust. However, the normalized DTW and LTW outperform it.
After detecting the onset and completion of artifacts, the performance of LTW was better in {\it Inter-Session}. Moreover, in the case of {\it Inter-Subject} setup, LTW outperforms normalized DTW with minimal neighbors, and normalized DTW surpassed LTW when the neighbors in the classifiers were increased.

\section*{Summary}

This chapter proposed an intelligent threshold-based detection method to identify the onset and completion of the artifacts in the EEG, LTW, and DTW distance-based methods for classifying artifacts. It was also evident that LTW and normalized DTW works best in the classification of artifacts. Further, the proposed models were found to be sound to analyze and classify the EEG artifacts of unseen subjects for three classes. In the subsequent chapters, we discuss developing a feasible interface for speech and motor challenged.

\chapter{Brain-Computer Interface using artifacts signatures in Electroencephalogram}
\label{chap:single}
\section{Introduction}
People with a speech impediment as a consequence of medical conditions like paralysis/cerebral palsy need personal assistant devices that understands and interpret their motor-disabled gestures to interact or communicate with the external environment. Assuming a person is capable of employing at least one muscle to interact, in the contemporary state of technology, muscle-based interfaces are favored, due to the sparse data transfer speed of Brain-Computer Interfaces (BCI) \citep{nicolas2012brain}. As discussed in the previous chapter, artifacts can be prominently recognized in the EEG signal and can be efficiently predicted using simple time series and time warping analysis.

In this chapter, we propose two working BCIs with varying numbers of electroencephalography (EEG) electrodes setup and a smartphone. The first proposed BCI is developed by employing an eye blink detector as an input mechanism. The second BCI uses eye blink and jaw movement as the input mechanism. A person may be capable of communicating with smartphone applications to get simple chores completed. In particular, mapping EEG artifact signal patterns to the keyboard activities can be employed to produce words and sentences. In order to improve the efficiency of the proposed BCIs, we use word completion models to decrease the number of artifacts required. The system communicates with a text to speech synthesis (TTS) system and outputs the speech corresponding to the word the user expected to deliver.

The remainder of the thesis is organized as follows. Section~\ref{sec:T9} briefs the conventional T9 keyboard. Section~\ref{sec:TrilWrd} outlines the Google trillion word corpus, which is used as the dictionary of the proposed BCIs. Section~\ref{sec:syn} describes the Android TTS system employed in the BCI. Section~\ref{sec:typesblink} briefs the types of eye blinks. The first proposed eye blinks based BCI is explained in the Section~\ref{sec:sinEEGBCI}. Section~\ref{sec:fourEEGBCI} discusses the second proposed BCI, which uses eye blinks and jaw movements.

\vspace{-13mm}
\section{T9 Keyboard}
\label{sec:T9}
T9 stands for Text on nine keys. T9 is a predictive text technology for mobile phones that comprise a 3x4 numeric keypad. The objective of the T9 keyboard is to make it simpler to record text information. It enables the vocabularies to be accumulated by a single keypress for each character, which is a tremendous advancement over the multi-tap method used in traditional mobile phone text insertion, in which several letters are associated with each key, and choosing one letter frequently demands multiple keypresses.

T9 consolidates the collections of characters on each phone key, including a fast-access dictionary of vocabularies, powered with Web 1T 5-gram Version 1, contributed by Google Inc \citep{brants2006web}. It looks up in the dictionary all vocabularies resembling the sequence of keypresses and place them by frequency of use. For example, in English, 4663 matches "good", "home", "gone", "hood", etc. Such sequences are identified as textonyms; "home" is considered as a textonym of "good." T9 is encoded to favor the word that to be the most common "textonym", such as "good" over "gone" or "home", "hand" over "game", or "bad" over "ace" or "cad". While the user inserts matching keypresses, along with vocabularies and stems, the system further provides word completions. On a mobile phone with a numeric keypad, every time a key is pressed, the algorithm yields the letters that are most suitable for the keys pressed at that time. For instance, to insert the term 'the', the user would press 8, followed by 4, and 3; the display would present 't' followed by 'th' then 'the.' Once the less-common term "Felix" is planned, despite entering 33549, the display exhibits 'E,' followed by 'De,' 'Del,' 'Deli,' and 'Felix.' This phenomenon is an illustration of letters switching while inserting the words. 

\section{Google Trillion Word Corpus}
\label{sec:TrilWrd}
Google Research uncovered a word n-gram model for a family of Research and Development projects, such as speech recognition, statistical machine translation, spelling correction, information extraction, entity detection, and so on. Usually, such models have been computed from training corpora comprising of several billion vocabularies, and Google has been adopting the immense potential from data centers and distributed processing to prepare more extended training corpora. Google scaled up the volume of their data by orders of magnitude rising in a training corpus of one trillion words from public web pages. The dataset holds 1,024,908,267,229 vocabularies of streaming text with 13,588,391 distinct vocabularies, after dropping vocabularies that occur less than 200 times. The distinct vocabularies were sorted based on the number of occurrences. Based on the textonyms, the best five vocabularies were used to build the dictionary for the proposed BCIs.

\section{Android Speech synthesis}
\label{sec:syn}

Text-to-Speech (TTS), also known as speech synthesis, in Android, is a powerful feature that can be used to supplement the Android applications. This extension will increase comfort, and the applications become more beneficial. The devices or applications employing TTS technology embrace a wide variety of fields such as education, screen readers, mobile technologies, disabilities, and communications. The proposed BCIs utilize Android Text To Speech synthesizer API 4 \citep{android2018api} for speech synthesis, and it is detailed in the following sections.

\section{Types of eye blink}
\label{sec:typesblink}
A single blink is defined by the closing and opening of the eyelid. It is a fundamental purpose of the eye that nourishes spread tears over and eliminate irritants from the facade of the cornea. There are three kinds of eye blinks, as discussed below.

\subsection*{Spontaneous blink}
Spontaneous blinking is performed without any external or visible stimuli and internal effort. This kind of eye blinking is carried in the pre-motor brain stem and occurs without intentional efforts. Humans blink their eyes about 15 times per minute.

\subsection*{Reflex blink}
A reflex blink happens in response to an outside inducement. A few of the examples for the inducements are contact with the cornea or objects that arrive quickly in front of the eye. A reflex blink occurs faster than a spontaneous blink. 

\subsection*{Voluntary blink}
Voluntary blink is a conscious eye blink, which has a larger amplitude than a reflex blink. Voluntary eye blink is used for building the BCI.

\section{Single Electode EEG BCI}
\label{sec:sinEEGBCI}
In this section, we discuss the eye blink artifact-based BCI using a single electrode electroencephalogram (EEG) device.

\subsection{Experiment Setup} In this section, the proposed BCI employs a Mindwave mobile+ \citep{mindwave2015} machine, with a Bluetooth interface for EEG data retrieval at a sampling rate of 512~Hz. The raw signal obtained from the device is quantized potential difference estimated between frontal and earlobe, as discussed in Section~\ref{sec:SEEG}.

\subsection{Preprocessing the signal} 
\label{sec:prep}
\begin{figure*}[!ht]
\begin{minipage}[h]{1.0\linewidth}
 \centering
        \includegraphics[scale=0.6]{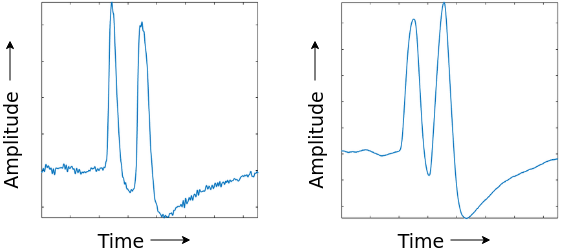}
  \caption{Raw EEG signal and corresponding moving average}
  \label{Fig:MovingAvg}
      \end{minipage}
\end{figure*}

Since the raw EEG signal is acquired using Bluetooth connection, noise may exist in the EEG signal. To eliminate the noise, the moving average ($\mathcal{M}_n$) is employed to smoothen the signal.
\begin{equation*}
\mathcal{M}_n = \frac{\Sigma_{i = n - m}^n t_i}{m}
\end{equation*}
where $t_i$ is the quantized EEG signal at the time instant $i$, $n$ is the current time instant, and $m$ is the number of samples in the EEG employed for moving average. Empirically we determined that $m=50$ operates best. An illustrative plot of raw and smoothed EEG signal is shown in the Figure~\ref{Fig:MovingAvg}.

\subsection{Personalized threshold} A small misplacement of the EEG electrode in the scalp may impact the robustness of the signal to a more substantial degree. The first twenty seconds are utilized to calibrate the personalized threshold for every subject.
\begin{equation*}
\textrm{Personalized\ Threshold}, P_t = \mu_{\mathcal{M}_n}  + (2 \times \sigma_{\mathcal{M}_n})
\end{equation*}
where $\mu_{\mathcal{M}_n}$ and  $\sigma_{\mathcal{M}_n}$ are the mean and standard deviation of moving average $\mathcal{M}_n$ respectively. Whenever the amplitude of the EEG signal crosses the estimated personalized threshold, the model predicts it as an occurrence of the eye blink.

\subsection{Eye blinks prediction} An empirical study of these samples reveals that a human eye blink is shorter than 500~ms, and the time pause between two voluntary eye blinks shorter than 1000~ms. The moving average ($\mathcal{M}_n$) for an interval of 1000ms is obtained. Once it passes the threshold, a timer thread is begun with a waiting interval of 1000~ms. Each time $\mathcal{M}_n$ passes the threshold, blink count is incremented, and the timer is reset to 1000~ms. Once the timer is expired, the number of eye blinks is delivered to the interface. The procedure for the eye blinks predication is given in Algorithm~\ref{algo:eyeblinks}.

\begin{algorithm}[H]
\setstretch{1.5}
\SetAlgoLined
\KwResult{noOfBlinks}
 noOfBlinks = 0\;
 timer = 1000 ms\;
 \While{timer not elapsed}{
  \If{$\mathcal{M}_n$ with positive slope crosses $P_t$}{
   noOfBlinks++\;
   Reset timer with 1000 ms\;
 }}
 \caption{Algorithm for eye blinks count}
 \label{algo:eyeblinks}
\end{algorithm}
Empirically, it is observed that two standard deviations from the mean are useful in predicting the eye blinks accurately.

\subsection{Virtual T9 Prediction Keyboard} 
\label{subsec:T9}

\begin{figure}[!htb]
    \centering
    \begin{minipage}{1\textwidth}
        \centering
        \includegraphics[scale=0.25]{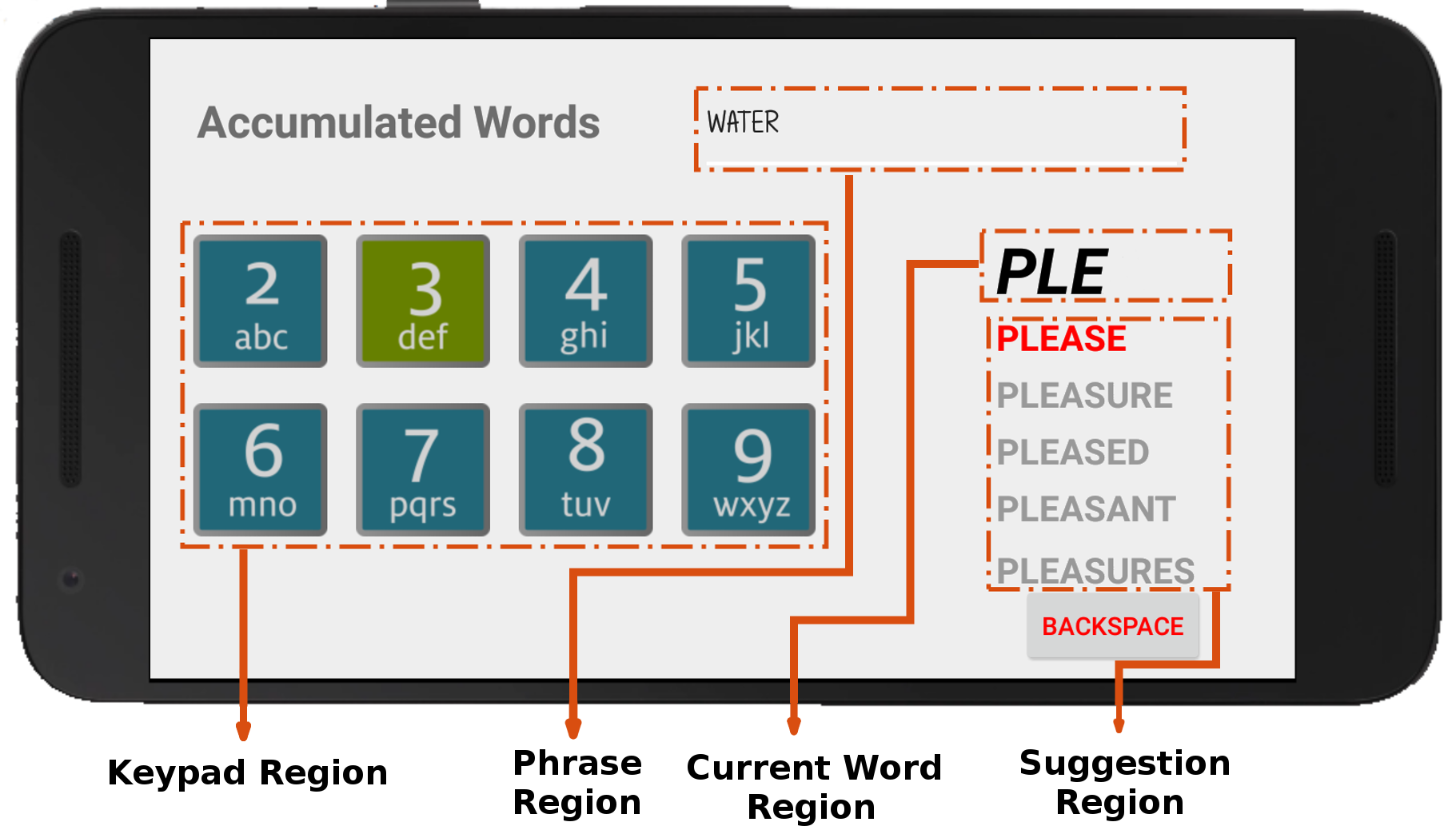}
  \caption{Screenshot of virtual T9 prediction keyboard}
  \label{fig:T9}
    \end{minipage}%
    \end{figure}

\begin{figure*}[!ht]
\begin{minipage}[h]{1.2\linewidth}
 \centering
        \includegraphics[scale=0.6]{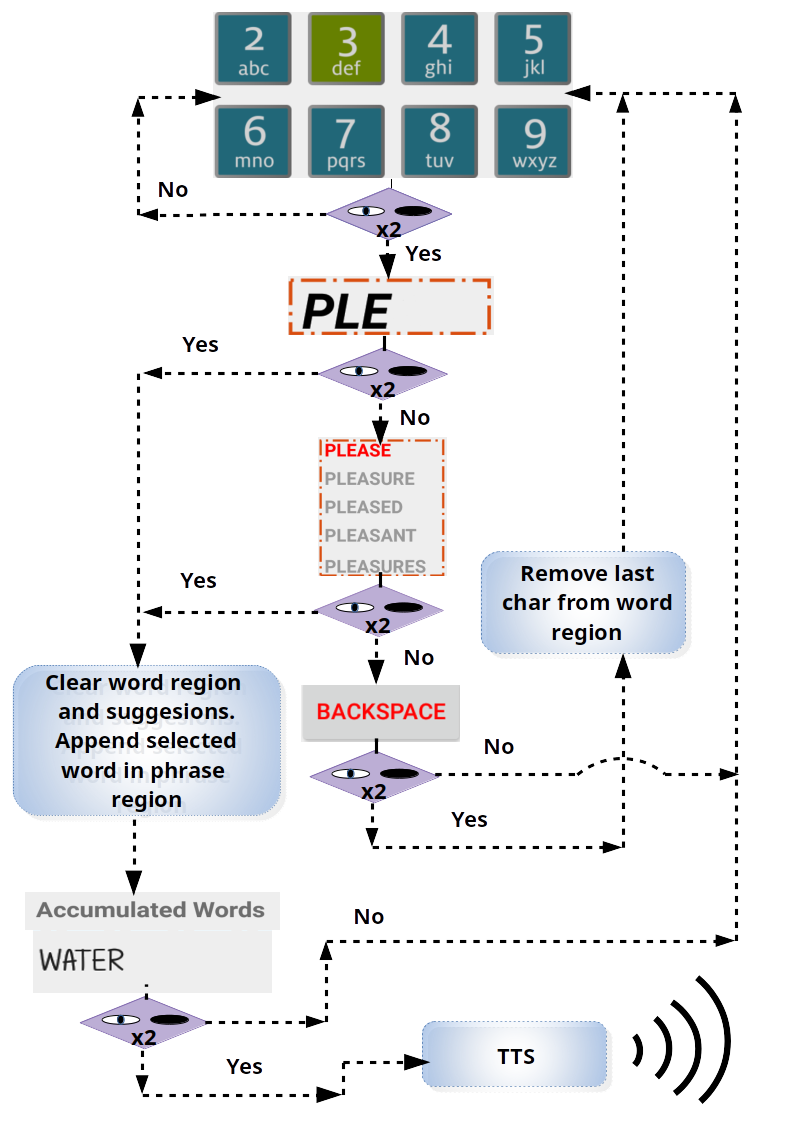}
  \caption{Flow chart of virtual prediction keyboard with eye blinks}
  \label{Fig:T9flow}
      \end{minipage}
\end{figure*}

T9 stands for Text on Nine keys, practices nine keys to serve the English alphabet, as in the Figure~\ref{fig:T9}. It does not need the multi-tap method employed in a conventional mobile phone keyboard. The virtual T9 keyboard is partitioned into four regions, namely keypad region, suggestion region, current word region, and phrase region. Fundamental navigation is aided by highlighting each key cyclically with a three seconds timer. The highlighting method starts from the keypad region and travels to different regions based on the user input eye blinks. In the keypad region, the user can pick the highlighted key by reacting with two voluntary eye blinks. Once a character is chosen, the character is appended to the current word region, and five suggestion words are presented in the suggestion region. These top five words are predicted using Google web trillion-word corpus, which was discussed in Section~\ref{sec:TrilWrd}. The words in the suggestion listing will further be highlighted following the other with a three seconds timer. Again the user can pick any one of the highlighted words from the suggestion listing with two voluntary eye blinks. The chosen words will be stored in the phrase region. The backspace button will be highlighted for three seconds. Blinking twice in that time will eliminate the last character in the current word region. Then the phrase region will be highlighted for three seconds. If the user makes two voluntary eye blinks in these three seconds, the words in the phrase region are sent to a TTS system, and the synthesized speech is played, as discussed in Section~\ref{sec:syn}. An illustrative flow chart for the process is shown in the Figure~\ref{Fig:T9flow}. This process happens infinitely till the application exits\footnote{A demo video is available in \href{https://github.com/harimaruthachalam/EyeBlinkBCI/blob/master/Demo/DemoBCIusingEEGofEyeBlinks.mp4?raw=true}{https://goo.gl/PSrsYe}.}.

\subsection{Virtual ABC Prediction Keyboard} 
\label{sec:ABC}
\begin{figure}[!htb]
    \centering
    \begin{minipage}{1\textwidth}
        \centering
        \includegraphics[scale=0.25]{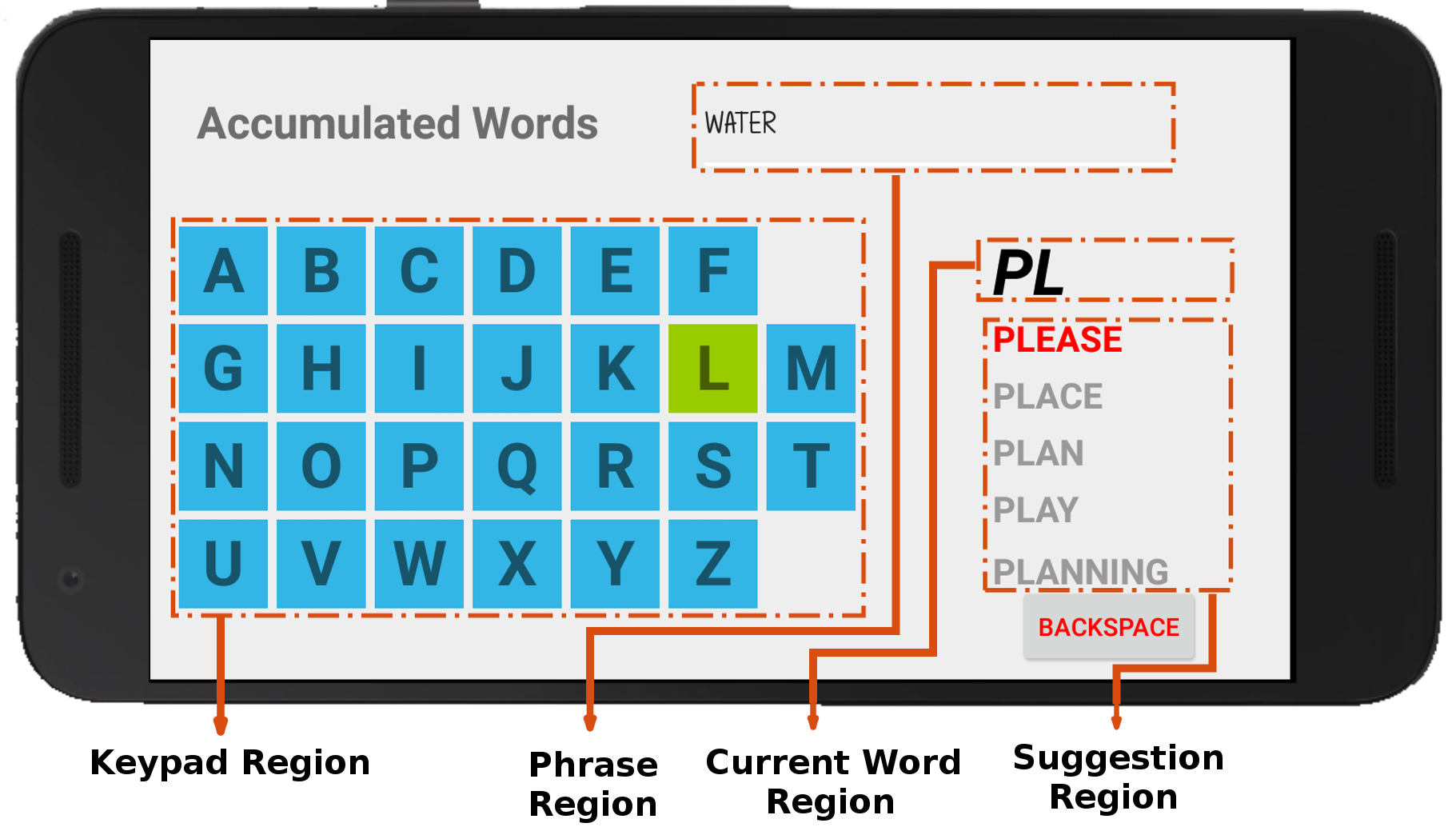}
  \caption{Screenshot of virtual ABC prediction keyboard}
  \label{fig:ABC}
    \end{minipage}
\end{figure}

The configuration of the virtual ABC keyboard, presented in Figure~\ref{fig:ABC} is comparable to the design of the virtual T9 keyboard. Virtual ABC keyboard includes of all the features as in the virtual T9 keyboard, accompanying with the convenience to speak nondictionary words, say names. The embodiment of processes of highlighting, typing, partial word completion model, navigating into suggestion list, picking a recommended word, affixing words to the phrase region, and presenting the TTS synthesized speech from the phrase region by eye blinks discussed in the subsection~\ref{subsec:T9}.

\section{Four Electrodes EEG BCI}
\label{sec:fourEEGBCI}
In the previous section, we discussed the proposed eye blink based BCI with a single electrode EEG. However, people with partial motor disabilities can make extended muscular movements, such as jaw movement. In this section, we propose extended and more effective BCI, which employs the eye blink and jaw movements artifacts.

\subsection{Experiment Setup and Preprocessing the signal} 

In this section, a four-channel muse EEG headset \citep{choosemuse} is used to acquire and stream the brain EEG signal, with a wireless Bluetooth connection to transfer the EEG signal to the Android mobile phone. A brief preface about the muse EEG headset was presented in subsection~\ref{sec:FEEG}. The acquired EEG signal was preprocessed, as discussed in subsection~\ref{sec:prep}.

\subsection{Artifact prediction}

The crucial part of building artifact-based BCI is to detect the presence of the artifacts in the EEG signal and predicting the type of artifact. A simple yet powerful threshold technique is used to detect the artifacts, and Dynamic time warping (DTW) is employed to classify the type of the artifact. The application is powered with a couple of artifacts for eye blink and jaw movement, which works as the reference sequence in DTW distance estimation.

\subsubsection{Detection}
The Android application is equipped with a template for eyeblink and jaw clench. The threshold for the artifacts is estimated from template sequences, as discussed in the Section~\ref{sec:thres}, with the difference that the number of electrodes in the muse EEG device being four. Whenever the realtime EEG crosses the threshold, we count the event as an appearance of the artifact. 

\begin{figure*}[!ht]
\begin{minipage}[h]{1.2\linewidth}
 \centering
        \includegraphics[scale=0.6]{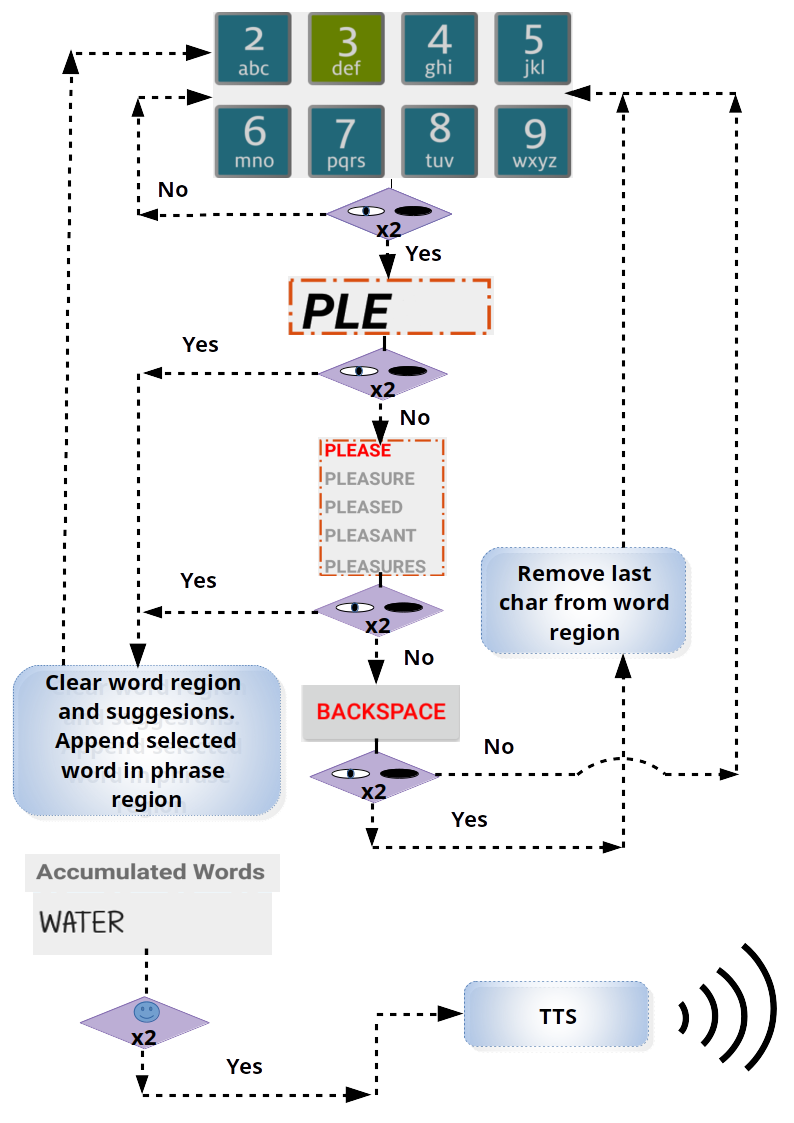}
  \caption{Flow chart of virtual prediction keyboard with eye blinks and jaw clenches}
  \label{Fig:chap5flow}
      \end{minipage}
\end{figure*}

\subsubsection{Classification}
Once the artifact is detected, a chunk of EEG signal with a duration of 500~ms starting from the detection of the artifact is extracted. The extracted EEG signal is classified to obtain the type of artifact using DTW based classification technique, as mentioned in the Section~\ref{sec:dtw}. Based on the type of artifact, the actions are mapped in the virtual keyboards, and it is discussed in the following sections.

\subsection{Virtual Prediction Keyboards}

The brief review about T9, ABC keyboards, their importance, developed T9, ABC virtual keyboards,  and its four regions, the navigation of each key, the operation of picking up the highlighted key with two consecutive eye blinks, character affixation, suggestion words mechanism, backspace mechanism, were discussed in subsections~\ref{subsec:T9} and \ref{sec:ABC}.

In contrast to the phrase region being highlighted for three seconds for two eye blinks, whenever the application encounters two consecutive jaw clenches, the words in the phrase region are sent to TTS, and the synthesized speech is presented. An illustrative flow chart for the mentioned process is shown in the Figure~\ref{Fig:chap5flow}. This process repeats infinitely until the application exits.

\section*{Summary}
In this chapter, we proposed two BCIs for people with speech or motor impediments. First, uses eyeblinks in the EEG signal, and whereas the second uses eyeblinks and jaw movement in the EEG signal to provide a series of English words that are sequentially synthesized to generate speech output. It is also evident from the chapter that the artifacts based BCI can be incorporated in various EEG apparatus.


\chapter{Summary and Conclusions}
\label{chap:sum}
Brain-Computer Interface (BCI) is a remarkable tool that understands brain signals. The system devised in this fashion can be useful for people paralyzed with spinal cord injury or Amyotrophic Lateral Sclerosis (ALS). The system enables the management of a computer or any other electronic device as well as to produce viable means of communication. The thesis deals with the online and offline detection, and classification of artifacts in Electroencelography (EEG) signals to develop and enhance the resolution of the new and existing BCI systems.

\section{Summary}
\label{sec:summary}
In the thesis, there are two primal challenges which have been tackled. The first is to detect the artifacts and predict the type of artifacts that appears in the EEG signal. A simple and more effective way to detect the artifacts is employed using a threshold based method. It is evident from Section~\ref{sec:det} that the proposed threshold method is more reliable in terms of detecting the appearance of any artifacts. Once the artifacts have been detected, the time warping techniques are used to classify the artifacts into various types based on its nature. The proposed linear and dynamic time warping techniques to classify the type of artifacts are useful in terms of classification accuracies.

The second aim is to utilize the proposed artifacts detection and classification techniques to build working, real-time, easy to use BCIs for people with partial or complete motor disabilities. A simple BCI with threshold-based eye blink detection was developed and reported with the aid of a single electrode EEG headset. Further, as an extension of this work, an efficient time warping classification based BCI was developed with a four electrodes EEG headset.

\section{Criticism}
\label{sec:criticism}
\begin{itemize}
\setstretch{1.5}
\item In chapter~\ref{chap:time}, in contrast to the conventional method of data split, which is 80\% of data for training or reference and 20\% of data for testing, the data split was 50\% for training or reference and 50\% for testing. There are two reasons for this. The first rationale is that the time warping techniques are of time complexity $O(n^2)$ for each reference, increasing the reference data would slow down the estimation of time warping distances. The second is that we used in the chapter is distance-based $k$-Nearest Neighbour classifier. $k$-Nearest Neighbour classifier needs all the reference data in the memory, which also leads to memory constraints.

\item How does it quantify to designate artifacts based speller as a Brain-Computer Interface (BCI), as artifacts are originated from noncerebral areas? The people with Amyotrophic lateral sclerosis (ALS) have partial or limited muscular movements. Furthermore, the last set of organs to paralyze is the facial muscles. So, it is sensible to augment the existing and proposing BCIs with the aid of the artifacts.

\item Since the thesis work is on minimal muscular movement or artifacts based BCI, why don't we use a camcorder to record the streaming video to detect and classify the artifacts? Artifacts like eye blinks are shorter in nature, hardly 500~ms. The data acquisition system used in the proposed model uses a sampling rate of 512~Hz, whereas the framerate of a camcorder is relatively way less. Moreover, artifacts impact the EEG signal to a more significant extent; so, it is comparatively simple to detect and classify the artifacts in the EEG signal.

\item In chapter~\ref{chap:single}, the proposed BCIs uses artifacts detection and classification algorithm. However, the proposed algorithm assumes that no two artifacts co-occur. What happens to the proposed BCIs, if subject clenches the jaw and blink the eye simultaneously? The subjects are partially paralyzed and posses limited muscular movement and use the proposed BCIs to communicate to the outer world. Having said that, the subject seldom taunts the proposed models with simultaneous artifacts.

\item Dynamic Time Warping (DTW) is a conventional technique in signal processing and sequential pattern analysis. Moreover, DTW is a non-parametric pattern recognization methodology. Why can't we extend the proposed models with parametric techniques such as Hidden Markov Model (HMM) or connectionist models such as a Recurrent Neural Network (RNN)? From Section~\ref{Sec:Prep128EEG}, it is clear that the dimension of each trial is 128x750, and from the Table~\ref{Tab:TrialsCount}, it is apparent that we have 206 trials in total. Given the dimension and size of the trial, it is implausible to build a parametric model, as it leads to the curse of dimensionality. However, the counter-argument can be reducing the dimension of the trials. We have a collection of dimension reduction techniques; Principal Component Analysis (PCA) from statistics and Autoencoder from connectionist models. PCA demands the covariance matrix, which inturns require a substantial number of trials. Autoencoder necessitates training of the weight parameters, and we lack a considerable number of trials.


\end{itemize}

\section{Suggestions for future work}
\label{sec:suggestions}
\begin{itemize}
\setstretch{1.5}
\item In the thesis, it is evident that the threshold and the time-warping based techniques are robust for classifying eye blinks, head turn, head nod, and jaw movement. However, the study can be extended to various other facial muscle movements.
\item An extensive study on classification on various other artifacts can be aided in building a more sophisticated BCI.
\item In the thesis, the BCIs developed work robustly in English, whereas the interface can be extended to various other regional languages across the globe.

\end{itemize}

%
%
%


\begin{singlespace}
  \bibliography{refs}
\end{singlespace}


\listofpapers

\begin{enumerate}  
\item \textbf{Srihari Maruthachalam}, Sidharth Aggarwal, Mari Ganesh Kumar, Mriganka Sur, Hema A Murthy  \newblock
 \textbf{Brain-Computer Interface using Electroencephalogram Signatures of Eye Blinks.}
  \newblock {\em Interspeech 2018, Hyderabad, India}, pp. 1059-1060, 2018.
\item \textbf{Srihari Maruthachalam}, Mari Ganesh Kumar, Hema A Murthy  \newblock
 \textbf{Time Warping Solutions for Classifying Artifacts in EEG}
  \newblock {\em 41st Annual International Conference of the IEEE Engineering in Medicine and Biology Society (EMBC), Berlin, Germany}, pp. 4537-4540, 2019.
\end{enumerate}  

\clearpage
\leavevmode\thispagestyle{empty}\newpage

\begin{center}
\large{\bf{Graduate Test Committee}}
\end{center}

\begin{table}[h]
\centering
\begin{tabular}{lll}
\textbf{Chairperson} & \textbf{:} & Prof. N S Narayanaswamy                          \\
\textbf{}            & \textbf{}  & Department of Computer Science and Engineering \\
\textbf{}            & \textbf{}  & Indian Institute of Technology, Madras         \\
\textbf{}            & \textbf{}  &                                                \\
\textbf{Guide}       & \textbf{:} & Prof. Hema A. Murthy                             \\
\textbf{}            & \textbf{}  & Department of Computer Science and Engineering \\
\textbf{}            & \textbf{}  & Indian Institute of Technology, Madras         \\
\textbf{}            & \textbf{}  &                                                \\
\textbf{Members}     & \textbf{:} & Prof. Mitesh Khapra                              \\
                     &            & Department of Computer Science and Engineering\\
\textbf{}            & \textbf{}  & Indian Institute of Technology, Madras         \\
\textbf{}            & \textbf{}  &                                                \\
    &  & Prof. Srinivasa Chakravarthy                              \\
                     &            & Department of Biotechnology\\
\textbf{}            & \textbf{}  & Indian Institute of Technology, Madras    
\end{tabular}
\end{table}

\end{document}